\documentclass{aa}

\usepackage{graphicx}
\usepackage{amsmath,epsfig,rotating,amssymb}

\begin{document}

\def \rot{{\rm {\bf rot} }}
\def \grad{{\rm {\bf grad} }}
\def \div{{\rm div}}
\def \cha{\widehat}
\def \pr{{\it permanent}  regime }


\author{Hennebelle P. \inst{1}, Fromang S. \inst{2}}

\institute{Laboratoire de radioastronomie millim{\'e}trique, UMR 8112 du
CNRS, 
\newline {\'E}cole normale sup{\'e}rieure et Observatoire de Paris,
 24 rue Lhomond, 75231 Paris cedex 05, France 
\and  Department of Applied Mathematics and Theoretical Physics, University of Cambridge, Centre for Mathematical Sciences, Wilberforce Road, Cambridge CB3 0WA}

\offprints{  P. Hennebelle \\
{\it patrick.hennebelle@ens.fr}   }

\title{Magnetic processes in a collapsing dense core. I Accretion and Ejection}

\abstract
{It is important for the star formation process to understand the 
collapse of a prestellar dense core. }
{We investigate the effect of the magnetic field during the first collapse up 
to the formation of the first core,  focusing particularly
on the magnetic braking and the launching of outflows.}
{We perform 3D AMR high resolution numerical simulations of a magnetically supercritical 
 collapsing dense 
 core using the RAMSES MHD code and develop semi-analytical models that we
compare with the numerical results. }
{We study in detail the various profiles within the envelope of the collapsing core
for various magnetic field strengths. Even modest values of magnetic field strength
modify the collapse significantly. This is largely due to the amplification
of the radial and toroidal components of the magnetic field by the differential 
motions within the collapsing core. 
For a weak magnetic intensity corresponding to an initial mass-to-flux over 
critical mass-to-flux ratio, $\mu$ equals to 20, a centrifugally
supported disk 
forms. The strong differential rotation triggers the growth of a slowly 
expanding magnetic tower.  For a higher magnetic field strengths 
corresponding to $\mu=2$, the collapse occurs primarily along the 
field lines, therefore delivering weaker angular momentum in the inner 
part whereas at the same time, strong magnetic braking occurs.  
As a consequence no centrifugally supported 
disk forms. An outflow is launched from the central thermally supported core. 
Detailed comparisons with existing analytical predictions 
indicate that it is  magneto-centrifugally driven.  }
{For cores  having a mass-to-flux over critical mass-to-flux radio
 $\mu < 5$,  the magnetic field appears to have a 
significant impact. The  collapsing envelope is denser and flatter 
than in the hydrodynamical case and no centrifugally supported disk forms.
 For values $\mu < 20$, the  magnetic field drastically  modifies the disk 
evolution.  In a companion paper, the influence
 of the magnetic field on the dense core fragmentation is studied. } 

\maketitle

\section{Introduction}
Studying the collapse and the fragmentation of a protostellar molecular dense core is 
of great relevance for the star formation process. While  the role of the magnetic field
has long been suspected (e.g. Shu et al. 1987), it is still a disputable issue. 

The first calculations of a collapsing dense core  
were monodimensional and treated  ambipolar diffusion 
(e.g. Ciolek \& Mouschovias 1994, Basu \& Mouschovias 1995). 
Their main goal was to investigate the role of the magnetic 
support in delaying the protostar formation. At about the same time, a few attempts were 
made to calculate the collapse in 2 or even 3D 
(Phillips \& Monaghan 1985, Fiedler \& Mouschovias 1992).
In  parallel to the numerical efforts, various authors have looked for analytical solutions 
of this problem
(Galli \& Shu 1993ab, Nakamura et al. 1995, Li \& Shu 1996,  Basu 1997, 
Krasnopolsky \& K\"onigl 2002,  Hennebelle 2003, Tilley \& Pudritz 2003).

With the increasing computing power and the improvement of the numerical schemes, 
recent developments have   been realized and various 2D 
(Nakamura et al. 1995, Tomisaka 1998, Allen et al. 2003) as well as 3D numerical calculations 
have been performed (Hosking \& Whitworth 2004, Machida et al. 2005, 2007,  Ziegler 2005,
Banerjee \& Pudritz 2006, Fromang et al. 2006, Price \& Bate 2007). 

In these calculations, it has been found that the magnetic field plays a crucial 
role in the evolution of the collapsing dense core, in particular in the 
context of  fragmentation in multiple systems. 
It has also been found that outflows can be spontaneously 
launched during the collapse. These outflows have strong similarities with 
the one studied in many papers either numerically (e.g. Uchida \& Shibata 1985,
 Casse \& Keppens 2003, Pudritz et al. 2007) or analytically (e.g. Blandford \& Payne 1982, 
Pelletier \& Pudritz 1992, Contopulos \& Lovelace 1994, Ferreira 1997).

Here,  we present further 3D numerical calculations of a collapsing magnetised dense core. 
Our main goals are to investigate the influence of the magnetic field strength  on 
the collapsing envelope, the disk and the outflows. The fragmentation is studied in 
a companion paper (Hennebelle \& Teyssier 2007, hereafter paper II). 
In order to identify the physical mechanisms at play, we 
develop various analytical approaches that we then compare with the 
numerical solutions. 
The outline of the paper is as follow. In  the second part, the numerical setup 
and the initial conditions are presented. The third part studies the evolution of
the envelope. For this purpose, semi-analytical solutions are obtained and 
compared with the numerical results. The fourth part presents the results for the outflows. 
Comparison with classical analytical results are made.
In the fifth part, we  qualitatively compare our results with various observations, focussing 
particularly on the young class 0 source, IRAM04191 (Andr\'e et al. 1999, Belloche et al. 2002)
The sixth part concludes the paper.

\section{Numerical setup and initial conditions}

We perform 3D numerical simulations using the AMR code RAMSES (Teyssier 2002, 
Fromang et al. 2006). RAMSES is based on shock capturing schemes and can 
handle ideal MHD, self-gravity and cooling. It uses the constraint 
transport method to update the magnetic field and therefore is 
preserving the nullity of the divergence of the magnetic field. 
RAMSES has been widely tested and gives results 
comparable to other MHD codes for a large set of benchmarks. 
 The AMR scheme offers access to the high resolution needed to treat the problem. 
All the calculations performed in the following use the Roe 
solver. 

The calculations start with initially $64^3$ grid cells. As the collapse 
proceeds, new cells are introduced in order to ensure that the Jeans length
is described  everywhere with at least 10 cells. Nine levels of AMR are used
for a total of about 10$^6$  grid cells and an
  equivalent numerical resolution of $16384^3$.

Here, we consider simple initial conditions, namely an initially uniform sphere in solid 
body rotation. The magnetic field is initially uniform and parallel to the   
rotation axis. The sphere is embeded into a diffuse medium hundred times less dense.
This makes that the surface of the cloud is initially out of pressure equilibrium and therefore
expanding. However, since the cloud as a whole, is strongly self-gravitating, the collapse
is not affected.  
The motivation to start with such simple conditions, sometimes considered as 
the standard test case for gravitational collapse of dense cores, instead of, for example, 
with a quasi-equilibrium configuration, is twofold. First, the magnetised collapse has not been 
widely explored yet and we feel it is important at this stage to choose  conditions 
which can be easily reproduced by others. Second, unlike in the hydrodynamical case, 
when  magnetic field and rotation are considered, the age  of the structure does
influence the angular momentum distribution and the structure of the field lines.
This makes the choice of starting with such a structure in near equilibrium also questionable. 

Initially, the ratio of the thermal   over gravitational energy, $\alpha$, 
is about 0.37 whereas the ratio of rotational over gravitational energy, $\beta$, is equal to
0.045. These values are comparable to standard values quoted in various studies of dense cores and 
are not too far from typical  values inferred from observations. 
The cloud temperature is equal to 11 K. The cloud has a mass of one solar mass, a
radius of about $R_0 \simeq$0.016 pc,  a density $\simeq 5 \times 10^{-18}$ g cm$^{-3}$
giving a freefall time, $\tau_{ff} \simeq 3 \times 10^4$ years. 
In the companion paper (paper II), an $m=2$ perturbation of various 
amplitudes is added.  

The strength of the magnetic field is expressed in terms of mass-to-flux over 
critical mass-to-flux ratio, $\mu= (M/\Phi) / (M/\Phi)_c$, where 
$(M / \Phi)_c = c_1 / 3 \pi \times (5/G)^{1/2}$ (Mouschovias \& Spitzer 1976)
and $\Phi$ is the magnetic flux. $c_1$ has been estimated to be about 0.53.
The case $\mu=1$ corresponds to a cloud just
magnetically supported, i.e. magnetic forces balance gravitational forces. 
Various magnetic  strengths are considered in the following, namely, 
$\mu=1000$ (quasi-hydrodynamical case), $\mu=20$ (very supercritical cloud), $\mu=5$
and $\mu=2$ (highly magnetised super critical cloud).

In order to avoid the formation of a singularity and to mimic the  fact 
that at very high density, the dust becomes opaque and therefore the gas 
becomes nearly adiabatic, we use a barotropic equation of state:
$C_s^2 = (C_s^0)^2 \times (1 + (\rho/\rho_c)^{4/3})^{1/2}$, 
where   $C_s \simeq 0.2$ km/s is the sound speed and
 $\rho_c=10^{-13}$ g cm$^{-3}$. Note that Masunaga \& Inutsuka (2000) 
demonstrate that this is a good approximation for a one solar mass core.

However, with such an equation of state, the timestep in the central part 
of the cloud becomes so small that it is difficult to follow the
collapse during a long period of time. In order to avoid that problem, 
we have also performed complementary simulations with a critical density 
$\rho_c/10=10^{-14}$ g cm$^{-3}$.

\section{Envelope evolution}
In this section, we study the properties of the various fields in the 
collapsing envelope. We first present our notations and a simple 
semi-analytical approach which will be useful to understand  the simulation results.

\subsection{Analytical model}
Here, we develop a {\it phenomenological model} for the profiles of the various
fields near the equatorial plane. 
We stress that the main motivation 
in carrying out such analysis is to have models to interpret more accurately the complex
numerical results. More elaborate models have been developed 
(e.g. Galli \& Shu 1993ab, Li \& Shu 1996, Krasnopolsky \& K\"onigl 2002)
assuming mainly  self-similarity or equilibrium. Since both are restrictive assumptions
and given the complexity of the numerical results obtained in the following, it is 
 unclear to which extent these models could be used for the purpose of comparison
although they undoubtedly provide a sensible hint on the physical processes.

\subsubsection{Notation and assumptions}
We consider an initially uniform cloud of mass $M_0$, initial radius $R_0$, 
in solid body rotation with angular velocity $\omega_0$ and  threaded by a 
uniform magnetic field $B_z^0$ parallel to the z-axis.

In the following, we use standard Cartesian coordinates $(x,y,z)$
and cylindrical coordinates $(r,\theta,z)$ therefore
$r=\sqrt{x^2+y^2}$.

Let $h$ be the scale height of the cloud near the equator, 
we write (see e.g. Li \& Shu 1996): 
\begin{eqnarray}
h(r)  = a \times r, \nonumber \\
\rho(r) = {d C_s^2 \over 2 \pi G r^2}, 
\label{nota} \\
B_z(r) = { H_z C_s^2   \over \sqrt{G} r }, \nonumber 
\end{eqnarray}
where $\rho$ and  $B_z$ are the density and  
z-component of the magnetic field near the equator respectively.
In the following, it will be assumed that $a$, $d$ and $H_z$
depend weakly on $r$.
It is well known that such scaling is a reasonable approximation in the envelope
 during the class-0 phase in particular before the rarefaction wave launched at the formation
 of the protostar has propagated significantly (Shu 1977).

The structures of the radial and azimuthal components of the magnetic
field are a little 
more complex. It is well known that for symmetry reasons, $B_r$ and $B_\theta$
vanish in the equatorial plane, $z=0$. Their values increase with $z$ 
until  they reach their maximum, after which they decrease with $z$.
Since here we are interested only in the value near the equatorial plane, 
we write as Krasnopolsky \& K\"onigl (2002)    
\begin{eqnarray}
\label{Bnota}
B_r(r,z,t) = H_r(r,t) \times {z \over h(r)} {C_s^2 \over \sqrt{G} r}, \\
B_\theta(r,z,t) = H_\theta(r,t) \times {z \over h(r)}
 {C_s^2 \over \sqrt{G} r}. \nonumber 
\end{eqnarray}
These two expressions are valid until $z \simeq h$. At higher altitude, 
$B_r$ and $B_\theta$ decrease and tend toward their value outside the core 
which in the present simulations is zero. Therefore, it is expected
that the values of $|B_r|$ and $|B_\theta|$ at a given radius, $r$, are maximum at 
the altitude, $z \simeq h(r)$, ${\rm max} (B_r(r,z)) \simeq H_r C_s^2 / \sqrt{G} r$
and ${\rm max} (B_\theta(r,z)) \simeq H_\theta C_s^2 / \sqrt{G} r$.

Thus, in the following, it seems appropriate to display the quantities
${\rm max} (B_r(r,z)) / B_z(r,0) = H_r / H_z$ and 
${\rm max} (B_\theta(r,z)) / B_z(r,0) = H_\theta / H_z$.

\subsubsection{Axial and radial components of the magnetic field}
Since throughout this work, field freezing is assumed, the magnetic flux, $\Phi$,
is conserved within the cloud. Therefore:

\begin{eqnarray}
\Phi = \int B_z \times 2 \pi  r dr = B_z^0 \times \pi R_0^2 = 2 \pi H_z (C_s^2 / \sqrt{G}) R_c,
\label{flux_conservation}
\end{eqnarray}
where $R_c$ is the cloud radius at the current time whereas $R_0$ is the 
initial cloud radius. Thus we have: 
\begin{eqnarray}
H_z =   (\sqrt{G}/ C_s^2) \times B_z^0 R_0^2 / (2 R_c).
\label{flux_cons_res}
\end{eqnarray}
Note that in this expression the cloud radius $R_c$ is not known. With our choice 
of initial conditions, $R_c$ does not evolve much during the class-0 phase and we will assume 
$R_c \simeq R_0$ in the following. This leads to
\begin{eqnarray}
B_z(r)=\frac{B_z^0}{2}\frac{R_0}{r}
\end{eqnarray}

The $r$-component is less straightforward to obtain. 
Its growth  is due to the stretching of the field lines 
by the differential motions within the cloud.
In the case of a
thin and isopedic disk, Li \& Shu (1997) demonstrated that the magnetic flux 
and gravitational potential are proportional through the cloud allowing one
to compute all components of the magnetic field once the gravitational 
potential is known.
Krasnopolsky  \& K\"onigl (2002) have  assumed that $B_r$ is simply
proportional to the magnetic flux. 
Since the $B_r$ component 
appears  difficult to predict quantitatively, we  simply write 
\begin{eqnarray}
H_r = \eta H_z
\label{eta}
\end{eqnarray}
and the value of $\eta$ will be estimated from the simulation.

\subsubsection{Density field}
In order to estimate the density, we write axial and radial 
equilibrium conditions. Although the cloud is not exactly in equilibrium 
since it is collapsing, such assumptions lead nevertheless to reasonable 
estimates of the density as long as the collapse is not strongly triggered
(Shu 1977, Hennebelle et al. 2003). 

The equilibrium along the z-axis, neglecting the azimuthal component 
of the magnetic field and the tension term $B_z \partial_r B
_r$, is:
\begin{eqnarray}
-C_s^ 2 \partial _z \rho + \rho \partial_z \psi - \partial_z
\left( {B_r  ^2 \over 8 \pi} \right) = 0,
\label{z_equilibrium}
\end{eqnarray}
where $\psi$ is the gravitational potential.
Integrating once, we obtain 
(using $\partial_z^2 \psi \simeq - 4 \pi G \rho$):
\begin{eqnarray}
C_s^ 2 \rho +  {1 \over 8 \pi G} (\partial _z \psi) ^2  + {B_r  ^2 \over 8 \pi} = K(r),
\label{z_equilibrium}
\end{eqnarray}
where $K(r)$ is a function of $r$.
Evaluating $K$ at $z=0$ and at $z=h$, and using the expressions stated by Eqs.~(\ref{nota}) 
and~(\ref{flux_cons_res}), we get

\begin{eqnarray}
 d \simeq d^2 a^2 + {\eta^2 \over 4} H_z ^2, 
\label{z_equilibrium}
\end{eqnarray}
where we have also used the approximation $\partial _z \psi \simeq -4 \pi G \rho h$.

The equilibrium along the radial direction is (neglecting again the 
influence of $B_\theta$)
\begin{eqnarray}
-C_s^2 \partial_r \rho  + {1 \over 4 \pi}
(- B_z \partial _r B_z + B_z \partial _z B_r) +  \rho \partial_ r \psi \simeq 0.   
\label{r_equil0}
\end{eqnarray}

Thus we obtain, with Eqs.~(\ref{nota}) and~(\ref{Bnota})
\begin{eqnarray}
  d  + { H_z^2 \over 4 } (1 + \eta/a) \simeq   a d ^2, 
\label{r_equilibrium}
\end{eqnarray}
where the gravitational force $\partial_r \psi$ has been assumed 
to be $\partial_r \psi \simeq -G M / r^2$ 
with $M \simeq \int 2 \pi r \times 2 h(r) \rho  d r $. 

$H_z$ being known from Eq.~(\ref{flux_cons_res}), 
Eqs.~(\ref{z_equilibrium}) and~(\ref{r_equilibrium}) can be solved numerically 
once $\eta$ is estimated from the simulation, to provide  the values
of $d$ and $a$.
 For the case $H_z=0$, we have $a=d=1$, i.e.  the structure 
of the cloud is 
not modified by the magnetic field and therefore 
the density is the Singular Isothermal Sphere (SIS) density
(since the analytical model does not consider the effect of rotation).

\subsubsection{Azimuthal magnetic field and rotation}
\label{angular_modele}
The azimuthal component of the magnetic field, as well as the rotation are 
more difficult to obtain. In order to do so, we adopt a Lagrangian approach, i.e. 
we follow the fluid particle and compute its momentum and azimuthal magnetic field
along time. For this purpose, we simply use the fluid equations with density and poloidal field given as described above. To use dimensionless quantities, we define
\begin{eqnarray}
 \widetilde{r} = r / R_0, \widetilde{M} = M(r) /  M_0 , 
 \widetilde{t} = t \times \sqrt{G M_0 / R_0^3} 
\label{dimension}
\end{eqnarray}

To compute the position of the fluid particle, we simply write 
  (neglecting the thermal pressure)
\begin{eqnarray}
d _t V_r \simeq  - {G_{\rm eff}  M(r_0) \over r^2} + {V_\theta^2 \over r}  
\label{radial_vel}
\end{eqnarray}
with $V_r = d r / dt$. In this expression, $M(r_0)$ is the mass of the cloud 
within a radius $r_0$ and $G_{\rm eff}$ is the effective gravitational constant 
$G_{\rm eff} = G \times (1-1/\mu)$. 
It will be assumed that $M$ remains constant during the collapse, i.e. we do not 
consider any accretion  which may arise along the pole. 
Thus, we obtain
\begin{eqnarray}
d _ {\widetilde{t}} \widetilde{V}_r \simeq  -   
{\widetilde{M}(r_0) \over \widetilde{r}(\widetilde{t})^2} + {L(\widetilde{t})^2 \over \widetilde{r}(\widetilde{t}) ^3}  
\label{radial_vel_norm}
\end{eqnarray}
where $L = \widetilde{r} \widetilde{V}_\theta$ is the momentum of the fluid particle.

The momentum equation is
\begin{eqnarray}
d _t (r V_\theta) = {1 \over 4 \pi \rho} ( B_r \partial_r (r B_\theta) + r  B_z \partial_z B_\theta )
\label{orthoradial_vel}
\end{eqnarray}
Gathering Eqs.~(\ref{nota}),~(\ref{Bnota}) and~(\ref{dimension}), we get
\begin{eqnarray}
d _{\widetilde{t}} L \simeq   K \times {H_z  H_\theta(t) \over a d },  
\label{orthoradial_vel_norm}
\end{eqnarray}
where $K= C_s^2 R_0 / (2 G  M_0)$.

Finally, the induction equation together with Eqs.~(\ref{nota}) and~(\ref{Bnota})
 leads to 
\begin{eqnarray}
d _{\widetilde{t}}  H_\theta \simeq  - \eta { L(\widetilde{t}) H_z \over \widetilde{r}(\widetilde{t})^2 } 
\label{orthoradial_B}
\end{eqnarray}

Once $a$ and $d$ are known, 
Eqs.~(\ref{radial_vel_norm}),~(\ref{orthoradial_vel_norm}) and~(\ref{orthoradial_B})
can be integrated with time to obtain the particle momentum. In the following, we 
use these equations to directly compare with the numerical results. 



\begin{figure}
\begin{center}
\includegraphics[width=6cm]{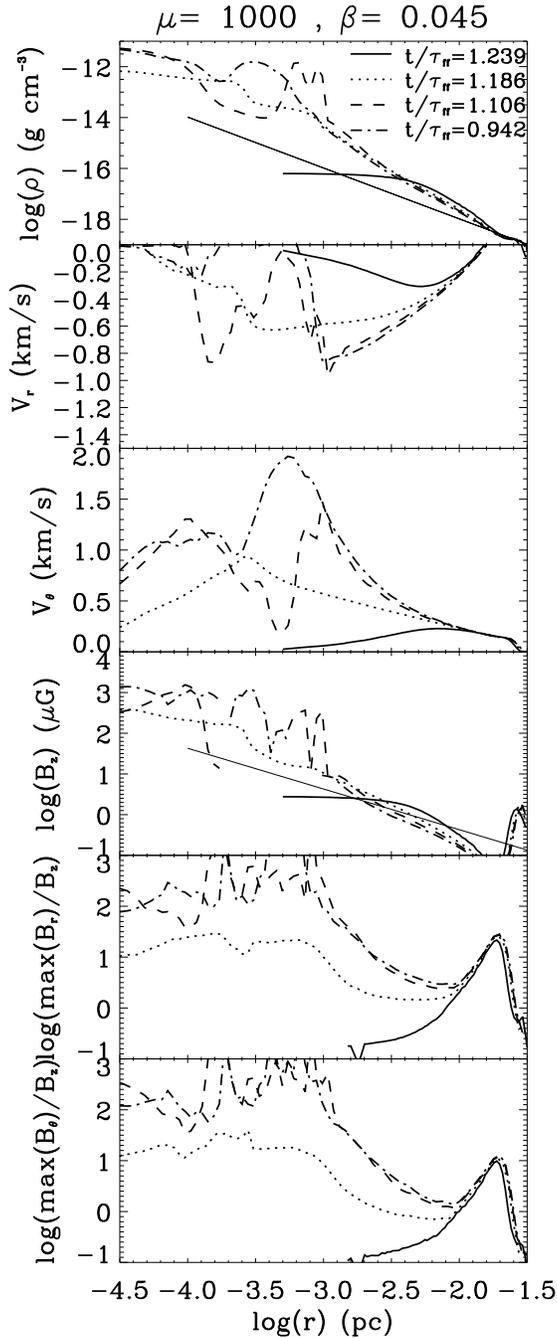}
\end{center}
\caption{Case $\mu=1000$. Density (left-top), radial (left-middle) and azimuthal velocity
(left-bottom) and
z-component of magnetic field (right-top) in the equatorial plane at different times.   
Largest values of the radial (right-middle) and azimuthal (right-bottom) magnetic components 
at a given radius are also given. For convenience, 
${\rm max}(B_r(r,z))/B_z(r,0)$ and
${\rm max}(B_\theta(r,z))/B_z(r,0)$ are given as a function 
of $r$. The various straight lines show analytical values (see text).}
\label{mu=1000}
\end{figure}

\begin{figure}
\begin{center}
\includegraphics[width=6cm]{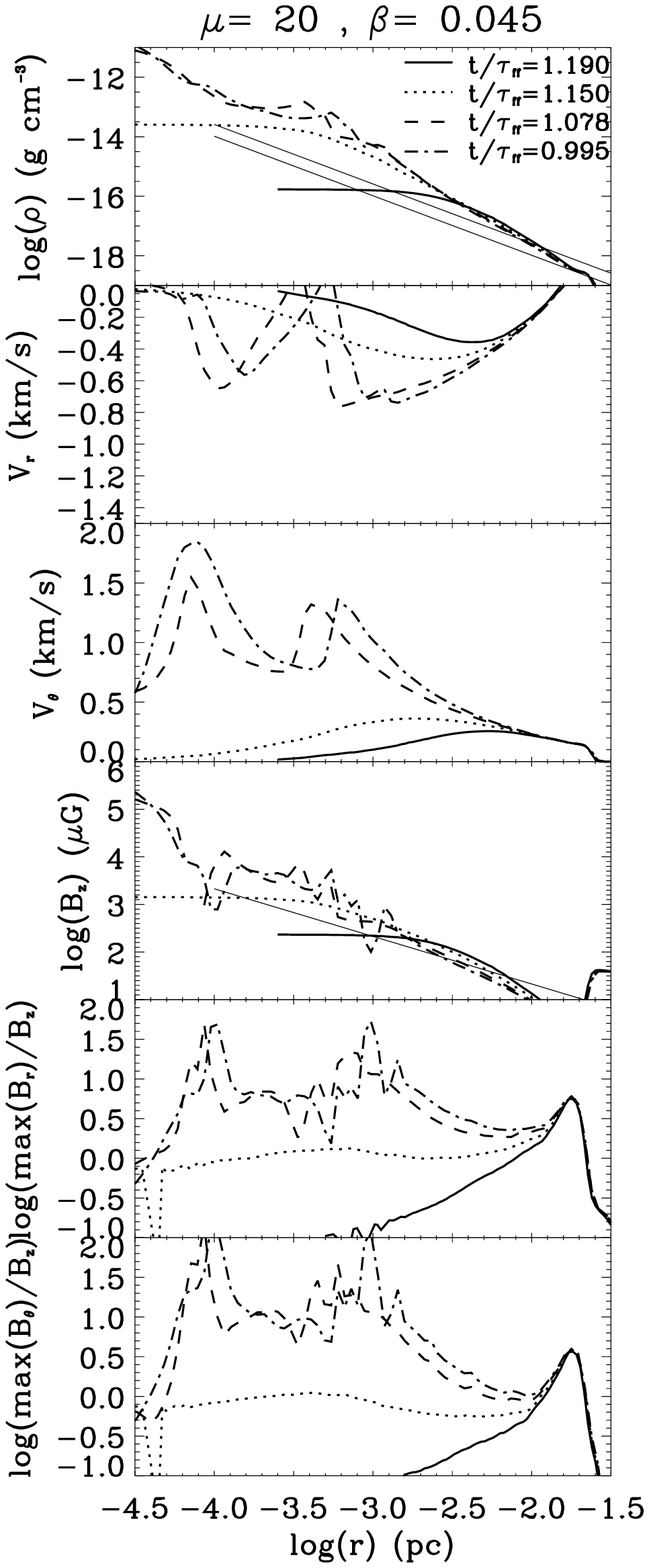}
\end{center}
\caption{Same as Fig.~\ref{mu=1000} for case $\mu=20$. }
\label{mu=20}
\end{figure}

\begin{figure}
\begin{center}
\includegraphics[width=6cm]{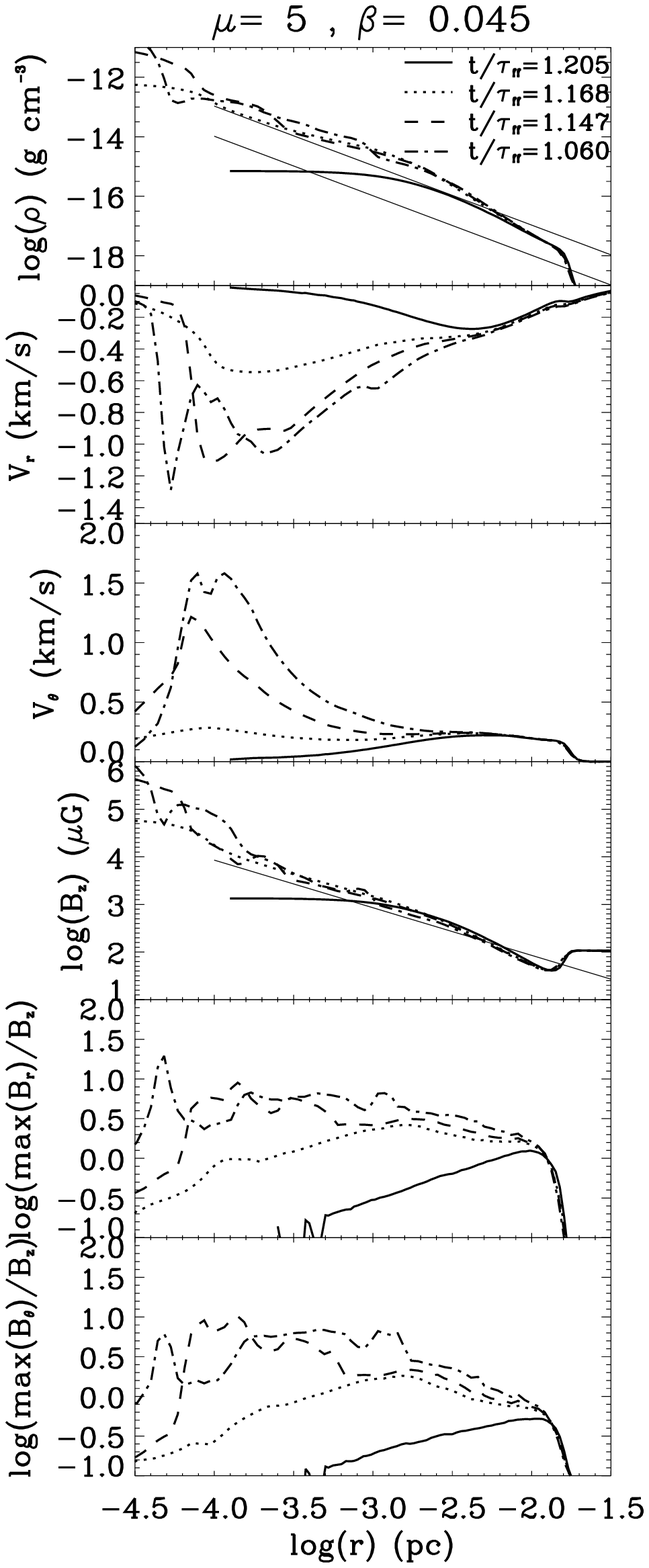}
\end{center}
\caption{Same as Fig.~\ref{mu=1000} for case $\mu=5$. }
\label{mu=5}
\end{figure}

\begin{figure}
\begin{center}
\includegraphics[width=6cm]{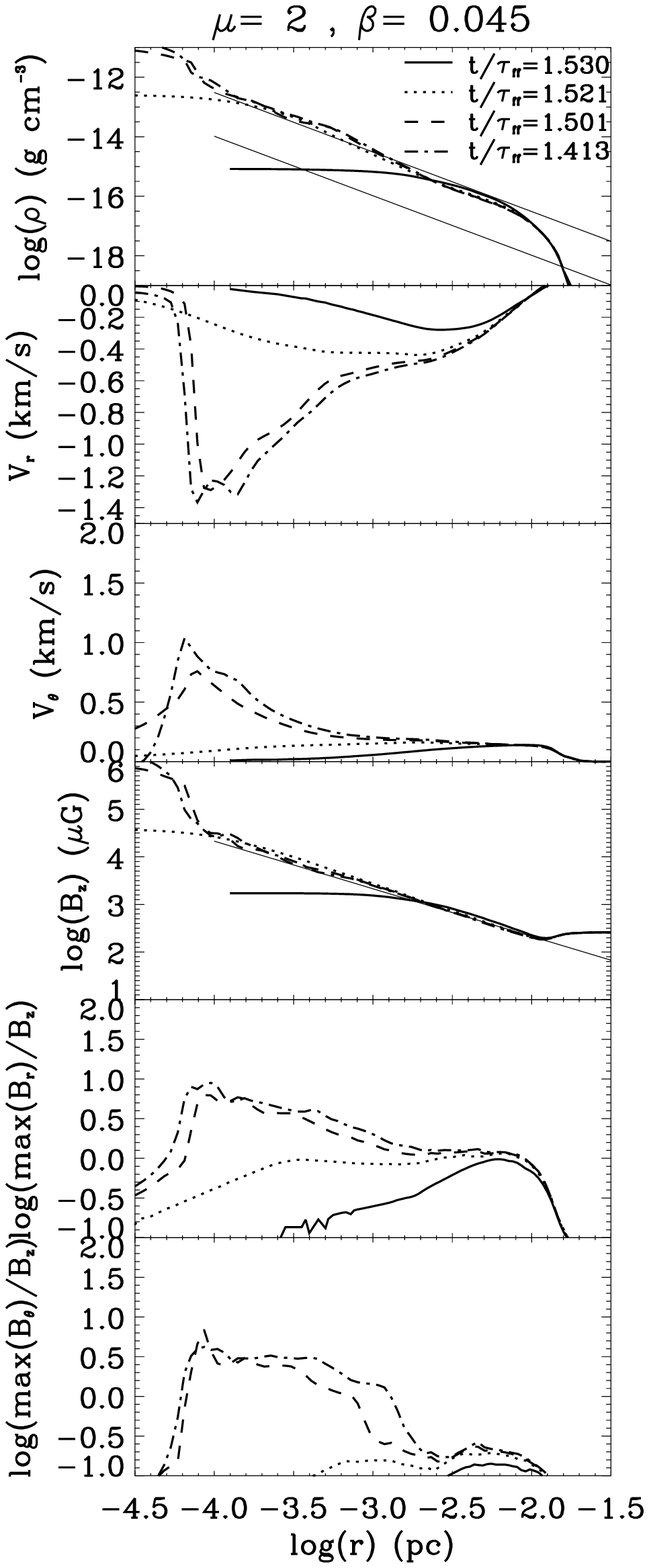}
\end{center}
\caption{Same as Fig.~\ref{mu=1000} for case $\mu=2$. }
\label{mu=2}
\end{figure}

\subsection{Cloud radial profiles}
Figures~\ref{mu=1000}-\ref{mu=2} show the density, radial velocity, 
rotation velocity and z-component of the  magnetic field in the 
equatorial plane (variations along the z-axis are shown in Sect.~4 ) 
as a function of radius for various magnetic field strengths. 
They also display the largest value
of radial and azimuthal components of the magnetic field at a given radius. These values
are obtained by taking the largest values along the z-axis at each radius. Note that 
as recalled in the previous section, $B_r$ and $B_\theta$ vanish in
the equatorial plane 
$z=0$. Therefore, the maximum value of $B_r(r,z)/B_z(r,0)$ at a given $r$ is plotted. 
Four snapshots are displayed. The first one is representative of the prestellar phase 
and is  about 0.06-0.08$\times \tau_{\rm ff}$ 
before density reaches the critical density, $\rho_c$, the second one is near 
the time at which the density reaches $\rho_c$ whereas the third and fourth ones
show latter evolution. 
The two straight solid lines in the density plots show the density of the singular isothermal sphere
(lower lines)
and the density of the analytical model stated by Eq.~(\ref{nota}) (upper lines).  
Note that in the hydrodynamical case, the two straight lines are indistinguishable.
Table~1 
gives the values of the parameters, $\mu$, $H_z$, $\eta$, $a$ and $d$.
The straight solid lines in the $B_z$ plots show the 
analytical estimate of $B_z$ stated by Eq.~(\ref{nota})  and~(\ref{flux_cons_res}).
 
\begin{table}
\caption{Values for the parameters. $\eta$ is estimated from the numerical simulation. $H_z$, $a$ and $d$ are obtained  from the analytical model.}
\begin{center}\begin{tabular}{|r||c|c|c|c|}
\hline
$\mu$ & $H_z$ & $\eta$ &  $a$ & $d$  \\
\hline
20 & 0.41 & 5 & 0.48 & 2.48 \\
\hline
5 & 1.64 & 3 & 0.2 &  10.29 \\
\hline
2 & 4.10 & 2 & 0.12 & 29.37 \\
\hline
\end{tabular}\end{center}
\end{table}

\subsubsection{Quasi-hydrodynamical case}
Figure~\ref{mu=1000} shows results for $\mu=1000$, i.e. quasi-hydrodynamical case. 
The density is slightly stiffer than $r^{-2}$ in the outer part where
it is a little denser than the SIS. This is due to the 
rotation and to the fact that the cloud is collapsing and therefore not 
in equilibrium (Shu 1977, Hennebelle et al. 2003, 2004a). In the inner part of 
the envelope the ratio of density over SIS density increases even more with radius.
This is due to the rotation velocity which increases with $r$ (Ulrich 1976). 
Note that a better 
agreement between analytical and numerical estimate can be obtained by taking into 
account the influence of rotation in the former (see e.g. Hennebelle et al. 2004a).
 Two accretion shocks are  visible in the radial velocity plot. 
The first one which is located at $r=10^{-3}$ pc shows the edge of the centrifugally 
supported disk. The second one,  located at  $r=10^{-4}$ pc, shows the edge
of the thermally supported core.
Although for this case, the magnetic field has almost no influence 
on the gas dynamics, it is worth studying the spatial dependence of the 
three components. The $B_z$ component in the envelope appears to be reasonably close 
to the analytical 
estimate stated by Eq.~(\ref{nota}), the discrepancy being due to the fact that 
$\rho$ is stiffer than $r^{-2}$ because of rotation. 
The $B_r$ and $B_\theta$ components  which vanish initially,
have slightly different behaviour. They grow with time 
and reach values  of the order of $B_z$ in the outer part of the 
envelope down to values roughly 10-100 times larger than $B_z$ at the edge of the disk.
Inside the centrifugally supported disk these values increase further up to 
values as high as about $10^3$. It should be noted however that here we are plotting 
the maximum values of $B_r$ and $B_\theta$ at a given radius. Since in the case $\mu=1000$,
the disk quickly fragments (see paper II), $B_r$ and $B_\theta$ fluctuate significantly
and therefore the high values reached in the inner parts are much higher than the mean 
values of $B_r$ and $B_\theta$ (see paper II for an estimate of their mean values in the disk).  
Note also that the increase of
$ {\rm max}(B_r) / B_z$ and ${\rm max}(B_\theta) / B_z$ at large radius ($r > 10^{-2}$ pc) 
is simply due to the decrease of $B_z$ in the external medium surrounding the cloud. 

\subsubsection{Weak field case}
Figure~\ref{mu=20} shows results for $\mu=20$, i.e. weakly magnetised case. 
As expected, since the magnetic field is weak, the density, radial and azimuthal
velocity fields are very close to those obtained in the previous case. 
$B_z$ is  much larger than in the case $\mu=1000$.
As for the previous case, the values
of ${\rm max}(B_r)/B_z$ and  ${\rm max}(B_\theta)/B_z$ 
 increase gradually in the envelope. They  
reach values of roughly  10 at the edge of the disk.  This indicates that the 
differential motions within the cloud are less important in this case 
because of  the influence of the Lorentz force. 
As in the hydrodynamical case, a centrifugally supported disk formed
as well as two accretion shocks.
Note again that the large fluctuations of $B_r$ and $B_\theta$ within the disk are due 
to the display of the maximum of $B_r$ and $B_\theta$ at a given radius. As shown in paper II,
some symmetry breaking is occurring in the disk which generates strong fluctuations. 

\subsubsection{Intermediate field cases}
Figures~\ref{mu=5} and~\ref{mu=2} respectively show results for $\mu=5$
and $\mu=2$, i.e. intermediate and strongly magnetised supercritical cases.
The density and velocity fields are significantly different from the two
preceding cases. The equatorial density is roughly 10 to 30 times 
the density of the SIS and is in good agreement with the analytical 
estimate stated by  Eq.~(\ref{nota}). This is mainly due to the
magnetic pressure (due to $B_r$) which compresses the gas along the z-axis. 
In the outer part, the radial  velocities
are  smaller than in the weak field cases.  
This is due  to the influence of the Lorentz force which supports the cloud. 
On the contrary, in the inner part, the radial velocities are larger than 
in the weak field cases. This is because, since the rotation is much 
smaller than in the weak field case, the centrifugal support is much 
weaker.
In the case $\mu=5$, a weak local maximum, due to the centrifugal force is nevertheless
still present at $r \simeq 2-3 \times 10^{-4}$ pc. However, unlike in the cases
$\mu=1000$ and 20, the radial velocity does not vanish except in the center. This 
indicates that there is no real centrifugally supported disk in this case. 
 For $\mu=2$, only the shock on the thermally supported core remains,
indicating that the centrifugal force is not able to stop the gas. 
The reason for  lower angular momentum will be analyzed in the next section. 
We note that similar conclusions have been recently reached by 
Mellon \& Li (2007). 
It is also apparent that the shape of the rotation velocity is flatter 
when $\mu$ is smaller: the rotation curve stays 
roughly constant until much smaller radii. 

The z-component of the magnetic field is very close to the analytical estimate 
in the envelope of the core.
The value of ${\rm max}(B_r) / B_z$ is about 2 at the edge of the core for 
$\mu=5$ and about 1 for $\mu=2$. It gradually increases inwards and reaches
values about 2-3 times larger in the inner part. The values of ${\rm max}(B_\theta) / B_z$
are typically 1.5 to 2 times smaller than ${\rm max}(B_r) / B_z$ for $\mu=5$
and about 3 times for $\mu=2$. 

Altogether these results illustrate that even for low to intermediate values of the magnetic 
strength, magnetic field can have a drastic influence on the cloud evolution as well as 
on the disk formation. 
This is due to the fact that the radial and toroidal components of the magnetic field,
which vanish initially, are strongly amplified during the collapse by the differential 
motions. This makes that  the radial component $B_r$ does not  increase  linearly with
the initial magnetic field strength since the field is easier to stretch  when it is 
initially weaker.

It is worth 
stressing that such values of $\mu$ in the range $5-2$ are compatible with the more 
pessimistic estimates
derived from measurements of the magnetic intensity in the  dense cores
(Crutcher 1999, Crutcher \& Troland 2000, Crutcher et al. 2004).
Since we find that dense cores having $\mu$ smaller than $\simeq$5 are qualitatively 
different from the hydrodynamical cores, we conclude that magnetic fields are 
playing an important role in the collapse of dense cores and therefore in the star 
formation process. 

\begin{figure}
\includegraphics[width=8cm]{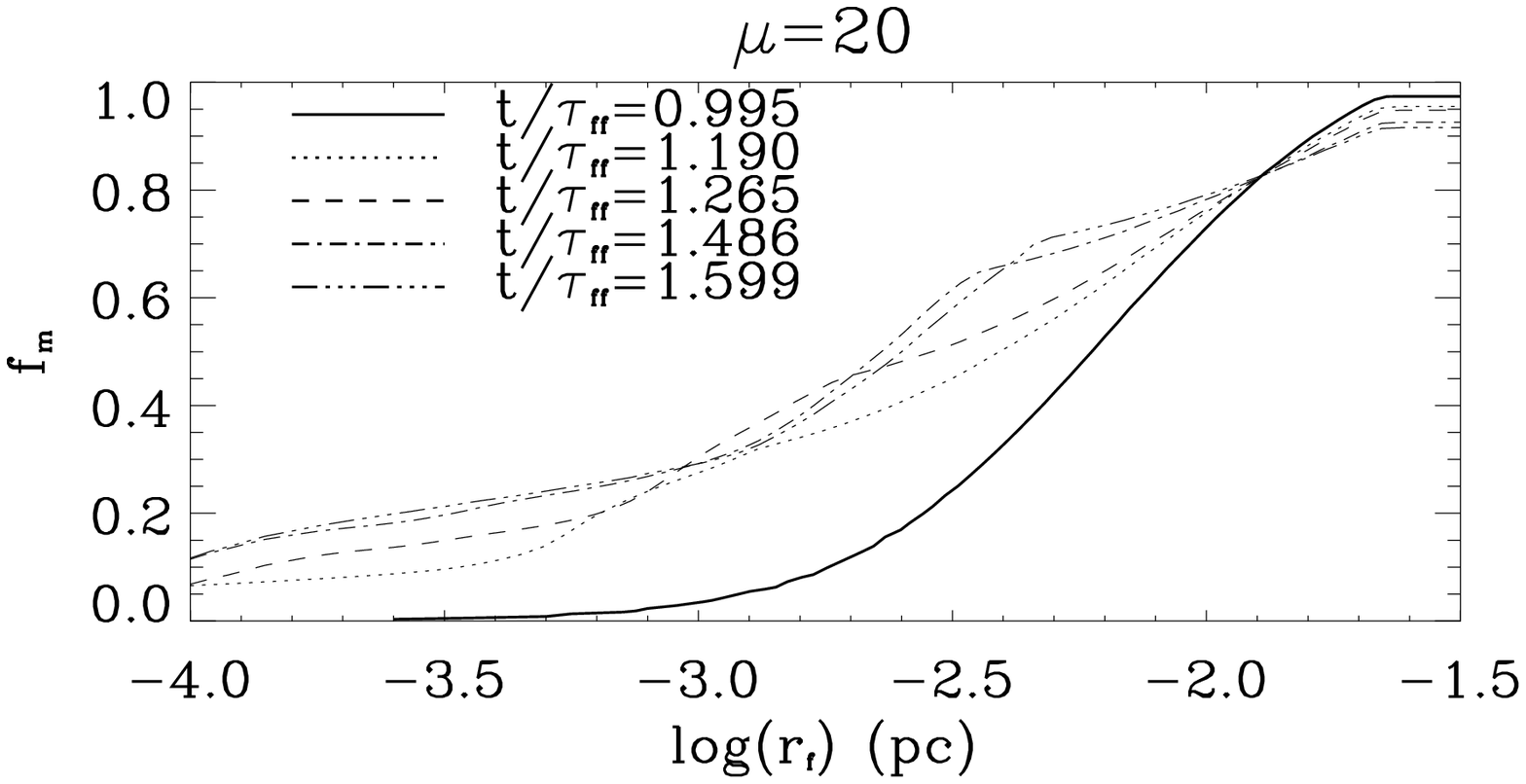}
\includegraphics[width=8cm]{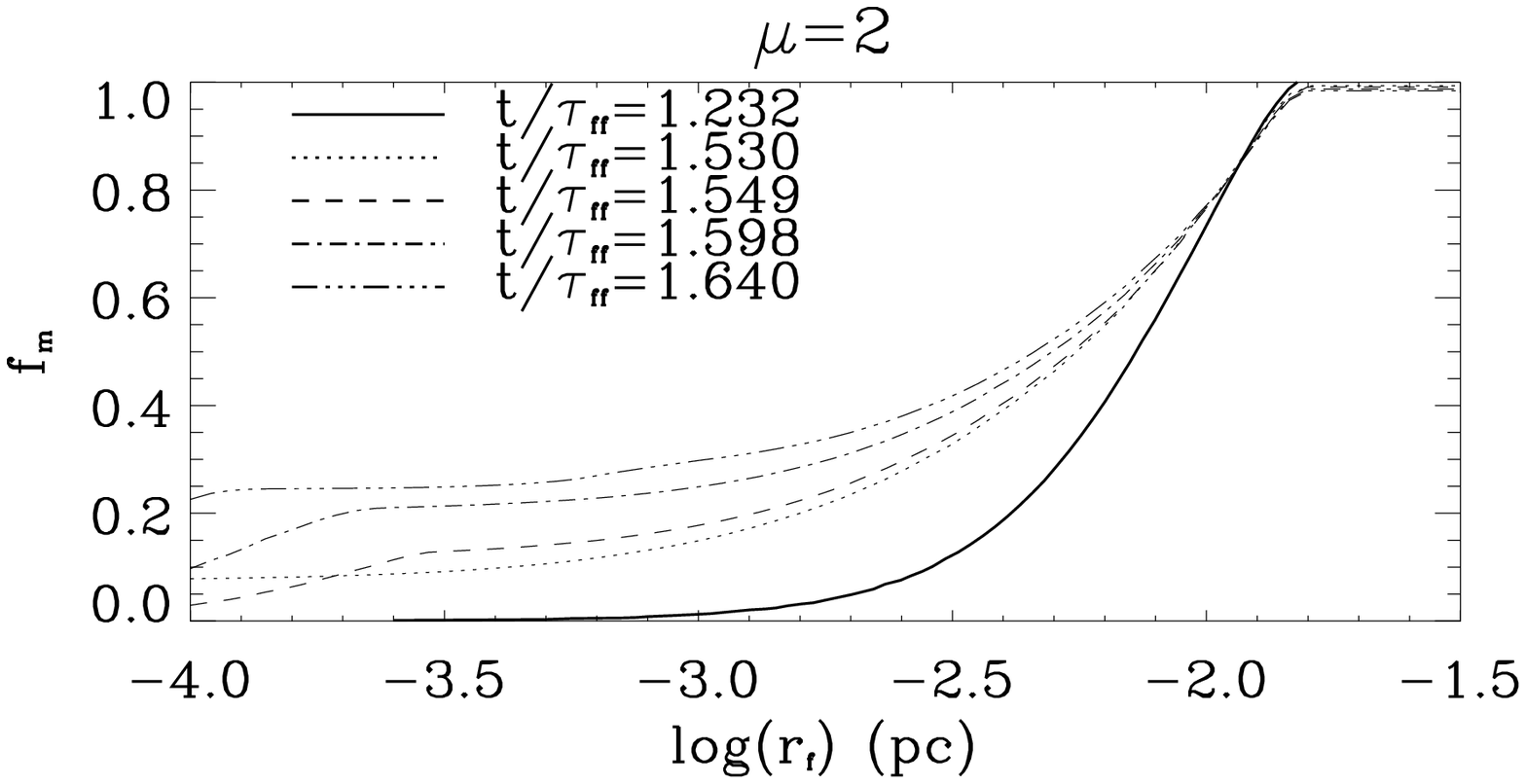}
\caption{Fraction of cloud mass, $f_m$, enclosed within a cylinder 
of radius, $r_f$, as a function of $r_f$.
Upper panel is case $\mu=20$. Lower panel is case $\mu=2$. }
\label{mom2}
\end{figure}


\begin{figure}
\includegraphics[width=8cm]{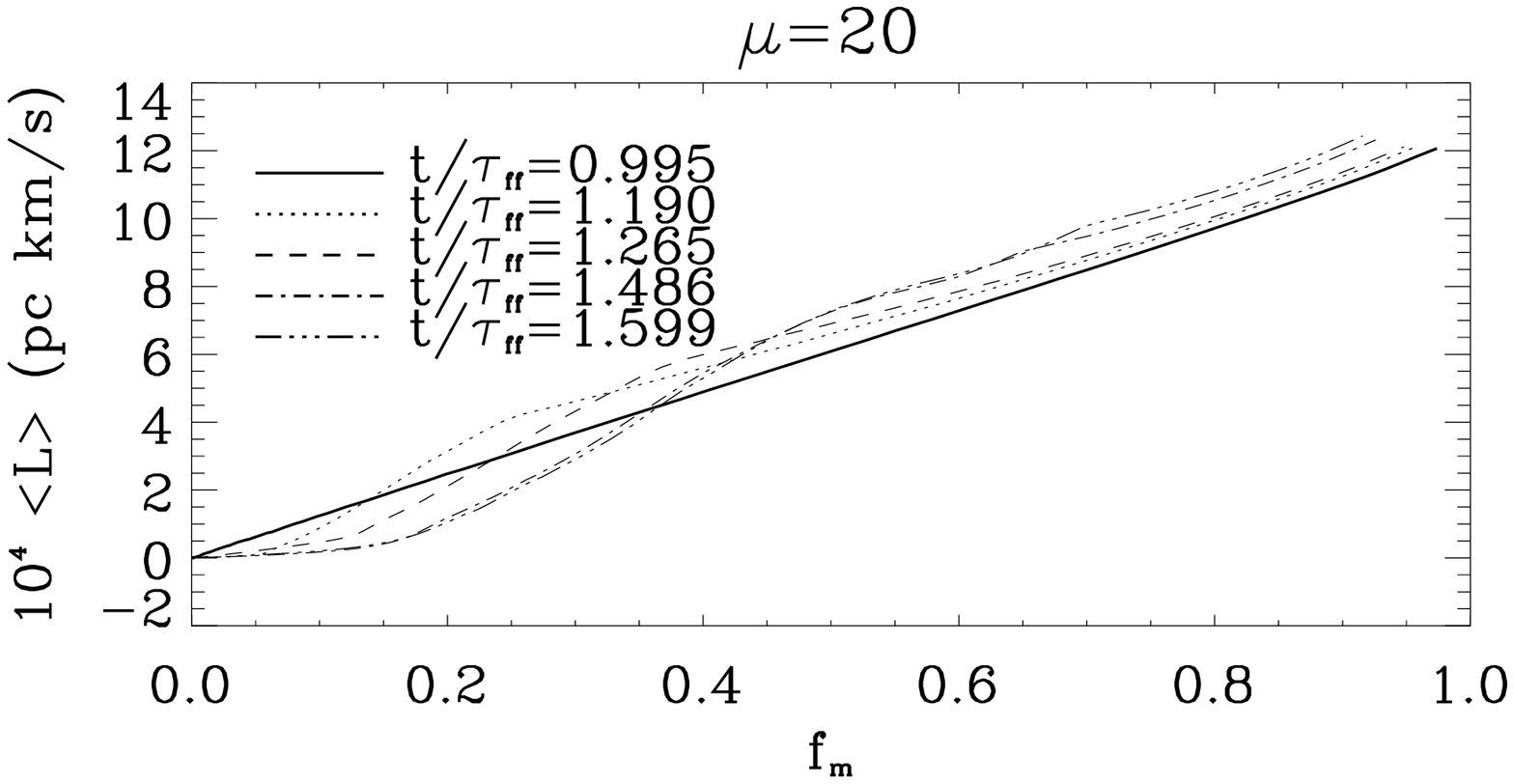}
\includegraphics[width=8cm]{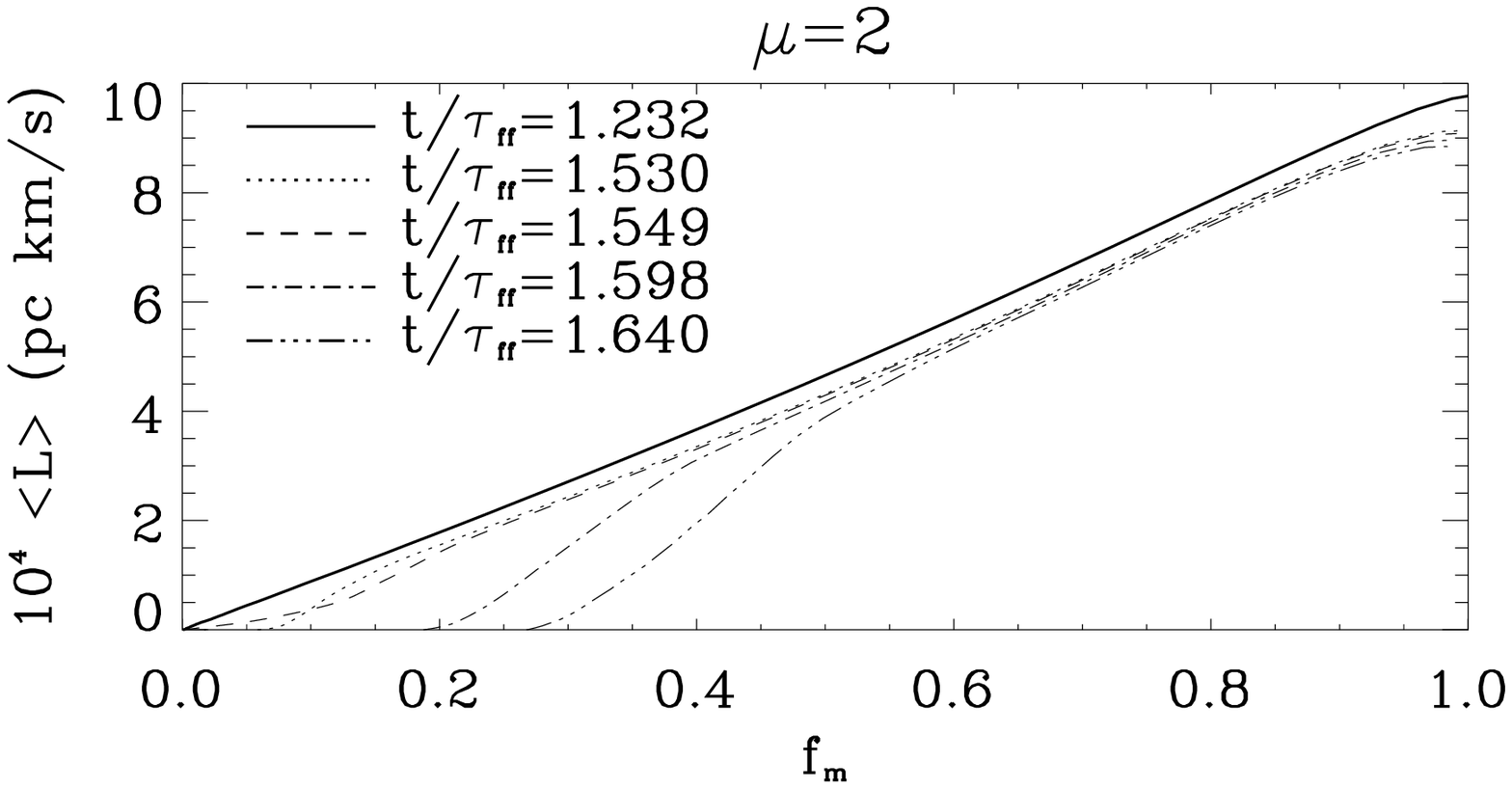}
\caption{Mean specific angular momentum enclosed within 
a cylinder of radius, $r_f$, containing a cloud mass fraction $f_m$,
 as a function of  $f_m$.
Upper panel is case $\mu=20$. Lower panel is case $\mu=2$. }
\label{mom1}
\end{figure}


\begin{figure}
\includegraphics[width=8cm]{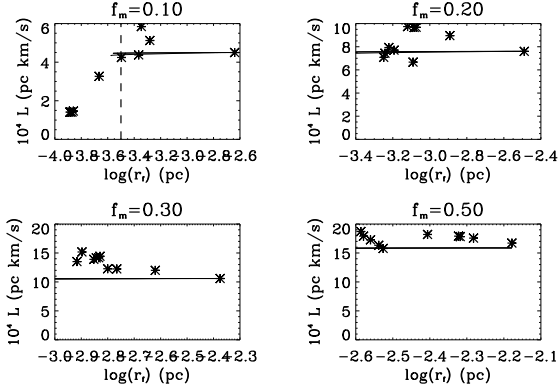}
\caption{Case $\mu=20$. 
Specific angular momentum of the fluid particle located at cloud radius, $r_f$,
where $r_f$ is the radius of the cylinder enclosing a constant mass fraction $f_m$.
Each point corresponding to a given time, this diagram shows the time evolution of 
the angular momentum of the  fluid particle as the collapse proceeds.
The solid lines correspond to the analytical model presented in Sect.~3 (see text). 
The dashed lines show the radius of the magnetized area which surrounds
the thermally supported core, below which the analytical model is not appropriate.}
\label{mom_evo_20}
\end{figure}

\begin{figure}
\includegraphics[width=8cm]{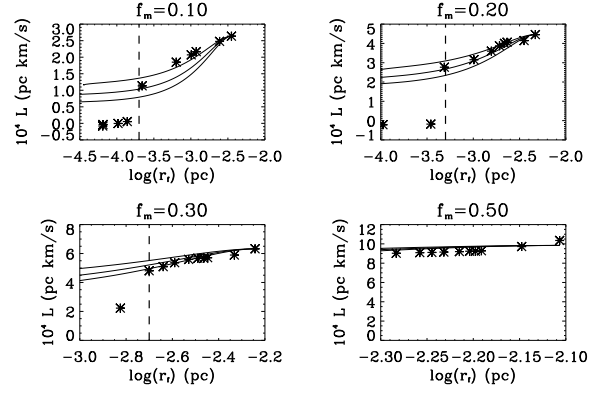}
\caption{Same as Fig.~\ref{mom_evo_20} for case $\mu=2$. }
\label{mom_evo_2}
\end{figure}

\subsection{Angular momentum evolution}
Here, we  further study the radial distribution of angular momentum. 
In particular, we investigate the physical origin of smaller rotation 
velocities in the intermediate and strong field cases. 

\subsubsection{Mass and angular momentum distribution}
For this purpose, we plot  the fraction of mass, $f_m$, enclosed inside 
cylinders of various radii for the cases $\mu=20$ and $\mu=2$ in  Fig.~\ref{mom2}.
Note that the first two times correspond to a critical density equal 
to $\rho_c$, whereas for the three others, the critical density is $\rho_c/10$. 
As can be seen, a good agreement is obtained between the second and the third times
which are close in time, showing that varying the critical density does not affect 
significantly the envelope evolution. 
 We define the radius $r_f$ of the cylinder containing a constant 
mass fraction $f_m$. $r_f$ decreases with time as the collapse proceeds.

Comparison between the 2 panels of Fig.~\ref{mom2}  reveals that the 
mass distribution  as a function of radius is significantly different 
in the two cases $\mu=20$ and $\mu=2$. In particular, the mass fraction  
enclosed within a cylinder of radius $\simeq 3 \times 10^{-3}$ pc is roughly 
fifty percent 
higher for $\mu=20$ than for $\mu=2$. This indicates that the collapse arises
in  different ways for these two cases. In fact, the collapse is more spherical 
for $\mu=20$ than for $\mu=2$. In the latter, since the field  is strong, the collapse
first proceeds along the field lines implying that the material  which constitutes 
the central core and the disk was originally located closer to the rotation axis 
than for the case $\mu=20$. Since material close to the rotation axis has a lower angular
momentum than the gas located further away, we believe that this is one of the 
reason of lower angular momentum in the cloud central parts in the case $\mu=2$
than in the case $\mu=20$.

Figure~\ref{mom1}  displays the distribution of mean specific angular
momentum within a cylinder enclosing a mass fraction $f_m$ as a function of $f_m$. 
 It shows that the specific angular momentum for the first two times are both
proportional to the mass fraction and that the cloud with $\mu=20$ has a specific angular
 momentum which is only 20$\%$ higher than for the case $\mu=2$. This difference 
which is attributable to the magnetic braking, shows that magnetic braking plays only
a minor role during the early parts of the collapse. Fig.~\ref{mom1} together with 
 Fig.~\ref{mom2}  also demonstrates that the total angular momentum 
in the case $\mu=20$ is higher in the internal part of the cloud  than in the case 
$\mu=2$. 
The subsequent times shown in these figures reveal  that the specific angular 
momentum stays roughly constant indicating that magnetic braking remains 
rather inefficient in the case $\mu=20$.  On the contrary, the subsequent times
displayed in the second panel of Fig.~\ref{mom2} show that the specific angular 
momentum decreases 
drastically in the inner part of the cloud. This is most likely due to the 
magnetic braking. We stress however, that at time $t=1.53 \tau_{ff}$,
the angular momentum has decreased significantly only 
in the very inner part corresponding to $f_m < 0.15$. With Fig.~\ref{mom1}, 
we see that this corresponds to radii smaller than $10^{-3}$ pc. Therefore,
the small rotation velocities seen in Fig.~\ref{mu=2} are largely due to the 
collapse proceeding first along the field lines. At this time, magnetic braking has
efficiently reduced the gas angular momentum  in the very inner parts only. 

\subsubsection{Comparison between analytical and numerical results}
In order to compare the numerical results with the analytical model and 
to confirm that the decrease of angular momentum seen for $\mu=2$ is due to magnetic 
braking, we have calculated the specific angular momentum 
of the fluid particle located at the radius, $r_f$, where 
$r_f$ is the radius of the cylinder which 
contains a cloud mass fraction $f_m$. Indeed, Fig.~\ref{mom1}
shows that the mean angular momentum enclosed within a cylinder of radius
$r_f$ does not vary much with time  (except for the 2 last times displayed for 
$\mu=2$). This implies that the mass enclosed within radius $r_f$, does not vary 
significantly along time. 
Therefore, the selected fluid particle located at $r_f$ should 
remain nearly the same. Consequently, any angular momentum variation 
is attributable to magnetic braking. Figures~\ref{mom_evo_20}
and~\ref{mom_evo_2} show, for different values of $f_m$, the 
specific angular momentum of the fluid particle 
as a function of radius at ten different times. 
It also displays 
analytical curves performed with the model presented in Sect.~\ref{angular_modele}.
To obtain these curves, we start with values of $r(t)$ and $L(t)$ corresponding
to the first point shown in each panel of Figs.~\ref{mom_evo_20} and~\ref{mom_evo_2},
and we integrate Eqs.~(\ref{radial_vel_norm}), (\ref{orthoradial_vel_norm}) and
(\ref{orthoradial_B}) using the values of $a$ and $d$ provided by 
Eqs.~(\ref{z_equilibrium})  and~(\ref{r_equilibrium}).
Note that in order to mimic the growth of $B_r$ and the fact that it is initially zero, 
we use $\eta = \eta_0 \times (1 - r(t) / r(0))$ in Eq.~(\ref{orthoradial_B}).
Since as shown in Fig.~\ref{mu=20}, the value of $\eta_0$ varies through the cloud, 
we run three models for $\eta_0$=1, 1.5 and 2. The top curves of Fig.~\ref{mom_evo_2}
correspond to $\eta_0=1$ whereas the bottom curves correspond to $\eta_0=2$.

The ten times represented in the case $\mu=20$ correspond to 
1, 1.15, 1.19, 1.2, 1.26, 1.35, 1.46, 1.54, 1.6, 1.63$\tau_{ff}$, 
whereas for $\mu=2$, they correspond to
 1.21, 1.41, 1.53, 1.52, 1.55, 1.57, 1.60, 1.62, 1.64, 1.68$\tau_{ff}$.
Note that for both cases, the first three times have been obtained 
with the standard critical density whereas the seven others have been 
obtained with the critical density $\rho_c/10$. The good continuity shows that 
the results are not affected by the thermally supported core (except maybe for  the 
last times).

In the case $\mu=20$, there is, as expected, hardly any variation of angular 
momentum. The only variation occurs for $f_m=0.1$ at  radius smaller than 
$r=3 \times 10^{-4}$ pc, i.e. after the fluid particle has reached the central 
thermally supported core. In the case $\mu=2$, magnetic braking is much 
more effective. A significant loss of angular momentum is observed for 
$f_m=0.1$, $f_m=0.2$ and $f_m=0.3$. In each case, the analytical fit is in reasonable
agreement with the numerical value until the fluid particle reaches 
a strongly magnetised area surrounding the thermally supported core
 where the analytical solution becomes inappropriate. This agreement shows that 
the analytical model  is reasonably accurate and that magnetic braking 
is responsible for the angular momentum decrease.
Depending on the value  of $f_m$, the angular momentum 
decrease during the collapsing phase can be larger, comparable or smaller than 
the angular momentum decrease once the fluid particle has reached the 
magnetised area which surrounds the thermally supported 
core. Note that the size of this area  increases with time 
due to accumulation of magnetic flux. 
This is why fluid particles corresponding to bigger $f_m$ 
reach it at larger radii. 
 
To summarize, we can say that for low magnetic strengths, magnetic braking is 
too small to play a significant role in the envelope. In the case of strong fields, 
the collapse first occurs along the field lines therefore delivering a low angular 
momentum in the inner region. At the same time, magnetic braking reduces the angular
momentum of the collapsing envelope. Finally,  strong magnetic 
braking occurs in the surrounding of the thermally supported core
which is highly magnetised.

\setlength{\unitlength}{1cm}
\begin{figure}
\begin{picture}(0,18)
\put(0,0){\includegraphics[width=6cm]{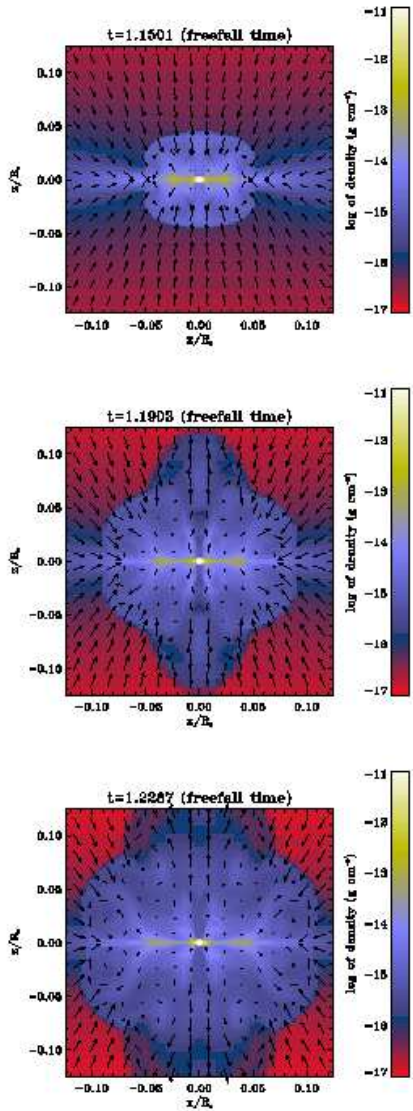}}
\end{picture}
\caption{Case $\mu=20$. Density and velocity fields in the xz plane.}
\label{xz_20}
\end{figure}

\begin{figure}
\begin{center}
\includegraphics[width=8cm]{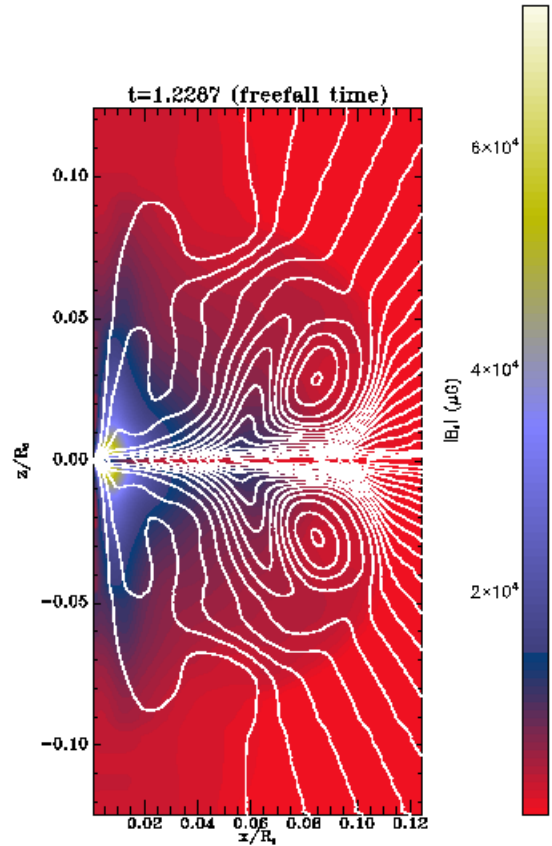}
\caption{Structure of the azimuthally averaged magnetic field in the
  model having $\mu=20$  at time $t=1.2287 \tau_{ff}$. The solid lines displays the
  poloidal magnetic field lines. They are overplotted on a snapshot of
  the toroidal magnetic field strength.}
\label{weakfield_lines}
\end{center}
\end{figure}

\section{Outflows}
It is now well known that accretion is often, maybe always, associated with 
ejection processes. In the context of star formation,   
molecular outflows  as well as jets have been extensively observed (see e.g. Bally et al. 2007
for a recent review). While jets may have velocities as large as several hundred km s$^{-1}$, 
the bulk of the millimeter wavelength CO emission, tends to have velocities of only a few to ten km s$^{-1}$.
In the following, we  call outflows, outward motions with velocities larger than 1 km s$^{-1}$.

In this section, we study the various outflow motions obtained
in these simulations. We first give a basic description of the weak and
strong field  cases. Then, a more detailed analysis is presented for
each of these two cases. Finally,  we show for both cases mass and
angular momentum fluxes for long time evolution.

\subsection{ Weak  field}
Here we describe results for the case $\mu=20$.

\subsubsection{Basic description}
Figure~\ref{xz_20}   shows the density and velocity fields 
in the xz plane for $\mu=20$ at three times. 
A complex expanding structure forms around the center.
As will be seen later, it is somehow similar to the magnetic tower 
investigated by Lynden-Bell (1996) and in the following 
we use this terminology.  
The first snapshot shows that this structure 
 encompasses the centrifugally supported disk. As a consequence 
the accretion shock, which occurs near the equatorial 
plane in the hydrodynamical case, is located further away at the edge
of the tower. 
At this stage, the tower is uniformly slowly  expanding  
(see next section for quantitative estimates).
The second snapshot shows that a faster outflow appears along 
the pole. It is clearly starting from the central thermally 
supported core. The velocity is all the way almost parallel to the 
z-axis.  Since this outflow is faster than the surrounding slowly 
expanding tower, the shape of the structure becomes gradually 
more complex and mainly composed of two distinct regions, 
the faster flow and a slower magnetic tower.
The third snapshot shows that this structure is maintained at later 
times without much change for the central flow  whereas the tower 
keeps expanding. At the edges of the structure, near the equatorial plane, 
 slow recirculation flows develop.

It should be noted at this stage that the thermal structure 
of the protostar is not correctly treated in this paper. In particular
the second  collapse is not considered here (Masunaga \& Inutsuka
2000, Machida et al. 2007). Thus the central outflow may have
a different structure in a more realistic simulation. Indeed, Banerjee
\& Pudritz (2006) and  Machida et al. (2006, 2007) found that a 
fast outflow, maybe a 
jet, having velocities around 30 km/s develops during the formation
of the protostar.

Figure~\ref{weakfield_lines} shows the structure of the magnetic field
at time $t=1.2287 \tau_{ff}$.
The magnetic field is decomposed into its 
toroidal and poloidal parts, $B_{\theta}$ and
${\bf{B_p}}=(B_r,0,B_z)$. The strength of the former is shown using
the colorscale snapshots while the poloidal magnetic field lines are
represented using the solid white lines.
The structure of the magnetic field appears to be complex. The field lines
are strongly bent and twisted in the inner central regions ($x/R_c< 0.1, \, 
|z/R_c|<0.1$) 
whereas they are almost straight in the outer part. 
In the same way, the toroidal component is 2 to 3 order of magnitude higher
 in the inner part than in the external part. This strongly suggests 
that the growth of the tower as well as the outflow are associated
with the strong wrapping of the field lines. This effect is quantified 
in the following section.

\begin{figure}
\begin{center}
\includegraphics[width=8cm]{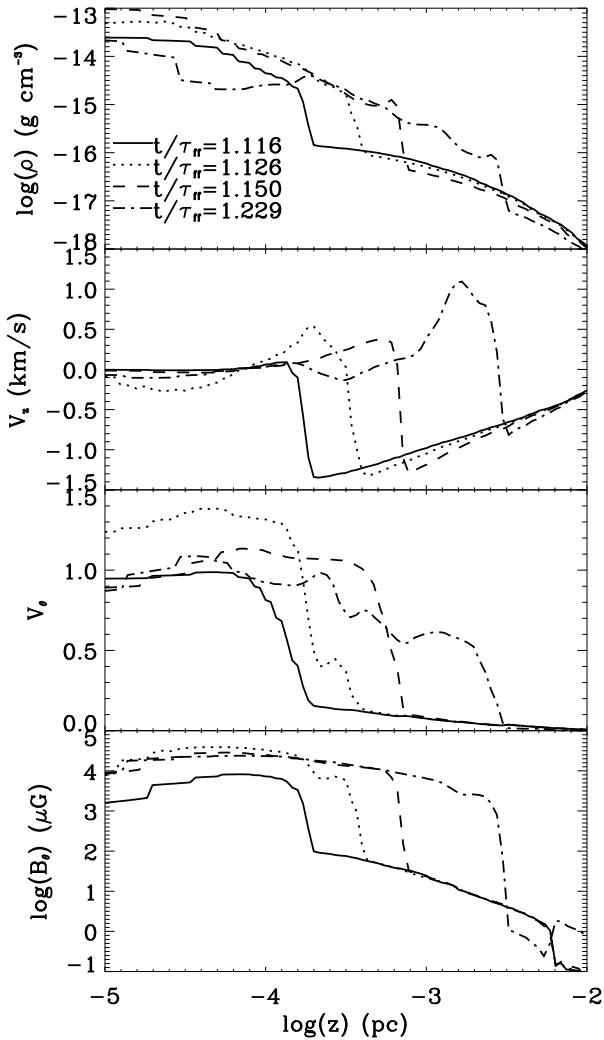}
\caption{Cut along the z-axis at $x/R_c=0.02$ and $y=0$
at four different times. Density, vertical component of the velocity, 
rotation velocity and toroidal component of the magnetic field
are shown. }
\label{magnetic tower}
\end{center}
\end{figure}

\subsubsection{Quantitative estimates}
Figure~\ref{magnetic tower} shows the density, axial 
velocity, rotation velocity and toroidal component of the magnetic 
field  along the z-axis for four times at $x/R_0=0.02$ and $y=0$.
The first and second times (respectively solid and dotted lines)
show that the central density is increasing due to the rapid accretion.
Similarly, the angular momentum increases. 
The toroidal component of the magnetic field grows rapidly and is about 5 times larger 
at the second time than at the first. This induces a slow 
expansion at about 0.3-0.5 km/s. 

The third time shows that the tower keeps expanding with about the same velocity
and that the toroidal component of the magnetic field does not grow in intensity 
and saturates, forming a plateau   (except close to $z=0$)
with  slowing decreasing  value  at high $z$.
The total toroidal magnetic flux inside the structure increases since the size
of this plateau increases.

To characterize the dynamical state of the tower, we estimate the thermal and magnetic
pressure as well as the gravitational potential at $z \simeq 3 \times 10^{-3}$ pc, i.e. 
close to the edge of the tower at time, $t= 1.15 \tau_{ff}$.
The density is about $10^{-15}$ g cm$^{-3}$, giving a thermal pressure of about 
$5 \times 10^{-7}$ erg cm$^{-3}$. The toroidal component of the magnetic field 
is about 10$^4 \mu G$ giving a magnetic pressure of about $B_\theta^2/8 \pi \simeq 
4 \times 10^{-6}$ erg cm$^{-3}$. The gravitational force is less straightforward to 
estimate. By the time we are considering, the mass denser than $\simeq 10^{-15}$ g
cm$^{-3}$ is of the order of $\simeq 0.1 M_s$. Thus, the potential energy is 
of the order of $\simeq \rho G M / z \simeq 10 ^{-5}$ erg cm$^{-3}$. 
Therefore, we conclude that by the time $t= 1.15 \tau_{ff}$, the magnetic tower is largely
 dominated by  the magnetic and the gravitational energies.
At later times, as the expansion proceeds, the gravitational energy
will eventually become negligible. 

To assess that the expansion of the tower is indeed due to the growth of the toroidal 
magnetic field, we consider pressure equilibrium at the edge of the tower
where we have 
\begin{eqnarray} 
B_\theta ^2 / 8 \pi = \rho_{\rm inf} V_{\rm inf} ^2,
\label{equil1}
\end{eqnarray} 
 since the 
external pressure is dominated by the ram pressure, $\rho_{\rm inf} V _{\rm inf}^2$, 
exerted at the accretion shock.

The flux per unit length of the toroidal magnetic field, $\Phi_\theta$, is given by 
\begin{eqnarray} 
\Phi_\theta = \int B_\theta dz \simeq B_\theta \times l, 
\end{eqnarray} 
$l$ being the height of the magnetic tower.

Integrating the induction equation along $z$,
$\Phi_\theta$ is also given by 
\begin{eqnarray} 
\Phi_\theta \simeq  B_z V_\theta \times t,
\end{eqnarray} 
where $B_z$ and $V_\theta$ are to be taken in the equatorial plane.
Therefore, we obtain:
\begin{eqnarray}
V_{\rm tower} \simeq {B_z  \over \sqrt{8 \pi \rho_{\rm inf} }} \times  {V_\theta \over V_{\rm inf}} .
\end{eqnarray} 
This expression is somehow similar to some of the expressions obtained by Lynden-Bell 
(1996, 2003) although his analysis is more  sophisticated since the
explicit value of the tower radius is taken into account. 
With $B_z \simeq 3 \times 10^3 \mu$G (obtained from Fig.~\ref{mu=20}),
$V_\theta \simeq 0.9 $ km/s, $V_{\rm inf} \simeq 1.3$ km/s and $\rho_{\rm inf} \simeq 3 \times 
10^{-17}$ g cm$^{-3}$ (obtained from Fig.~\ref{magnetic tower} either at 
$z=0$ or at $z=9 \times 10^{-4}$ pc), we obtain: $V_{\rm tower} \simeq 0.58 $ km/s.
This value is comparable with the value of $V_z \simeq 0.45$ km/s at time 
$t= 1.15 \tau_{ff}$ and $z=8 \times 10^{-4}$ pc within about $25 \%$.
The difference is probably due to the assumption of constant $B_\theta$ in 
the tower. Note that this simple estimate does not take into account gravity. 
In order to investigate its influence, an 
analytical model for the expansion of the magnetic tower,
 is developed in the Appendix. Indeed, the model shows that the growth of the 
transverse magnetic field which is induced by the gradient of transverse
velocity along the z-axis, triggers the expansion of a self-gravitating 
layer in a very similar way to what is observed in the simulation.

The last time in Fig.~\ref{magnetic tower} shows that the z-velocity increases significantly
 and reaches 
values of about 1.2 km/s. This is due to the central outflow  which  presents
higher velocities. At this stage the velocities of the tower and the flow are 
difficult to distinguish. 
The fourth time also reveals that  angular momentum as well as mass 
have been removed probably by the outflow
 between $z=3 \times 10^{-4}$ and $z=3 \times 10^{-3}$ pc. 

\subsection{Strong field: magneto-centrifugal ejection}
Here we describe results for the case $\mu=2$.

\setlength{\unitlength}{1cm}
\begin{figure}
\begin{picture}(0,18)
\put(0,0){\includegraphics[width=6cm]{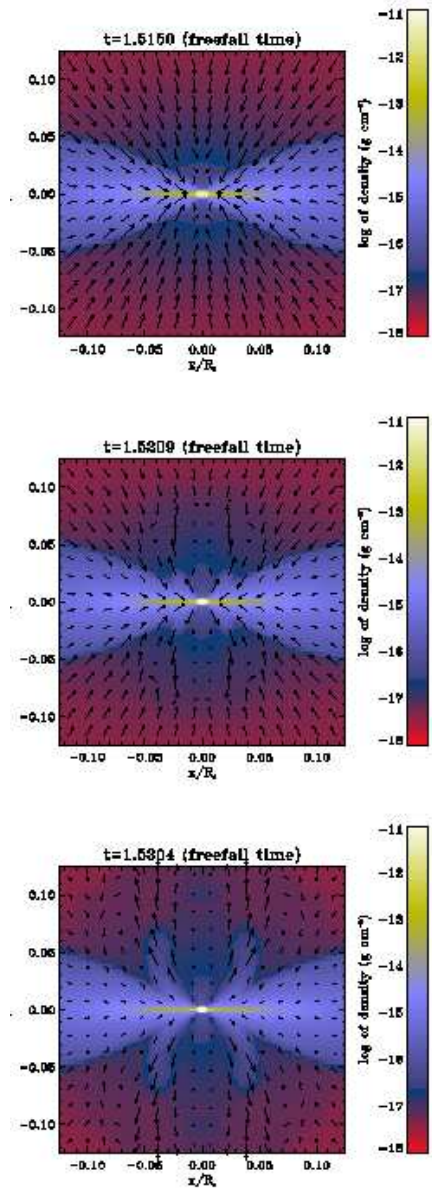}}
\end{picture}
\caption{Case $\mu=2$. Density and velocity fields in the xz plane.}
\label{xz_2}
\end{figure}

\begin{figure}
\begin{center}
\includegraphics[width=8cm]{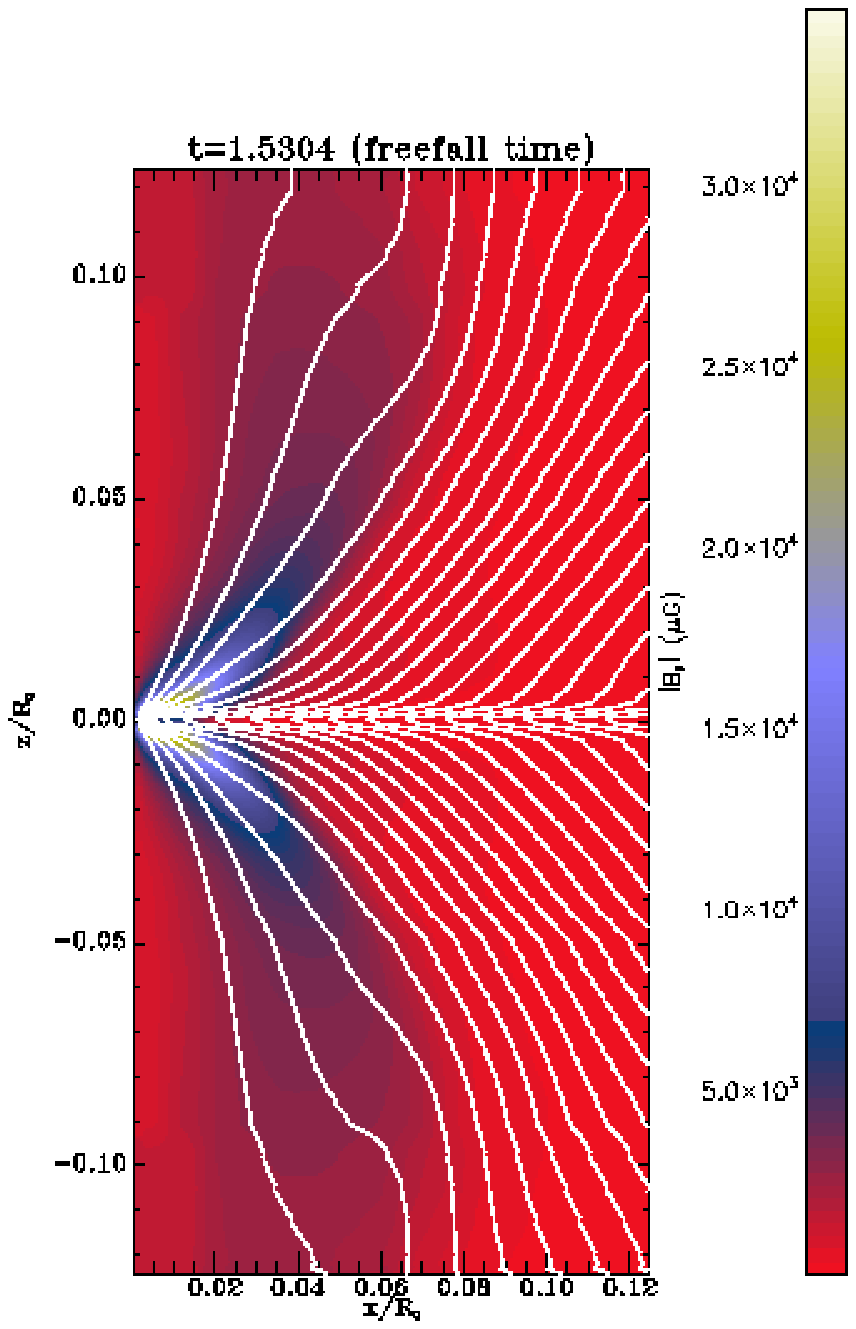}
\caption{Structure of the azimuthally averaged magnetic field in the
  model having $\mu=2$ 
  at time $t=1.5304 \tau_{ff}$. The solid lines displays the
  poloidal magnetic field lines. They are overplotted on a snapshot of
  the toroidal magnetic field strength.}
\label{strongfield_blines}
\end{center}
\end{figure}

\begin{figure}
\begin{center}
\includegraphics[width=8cm]{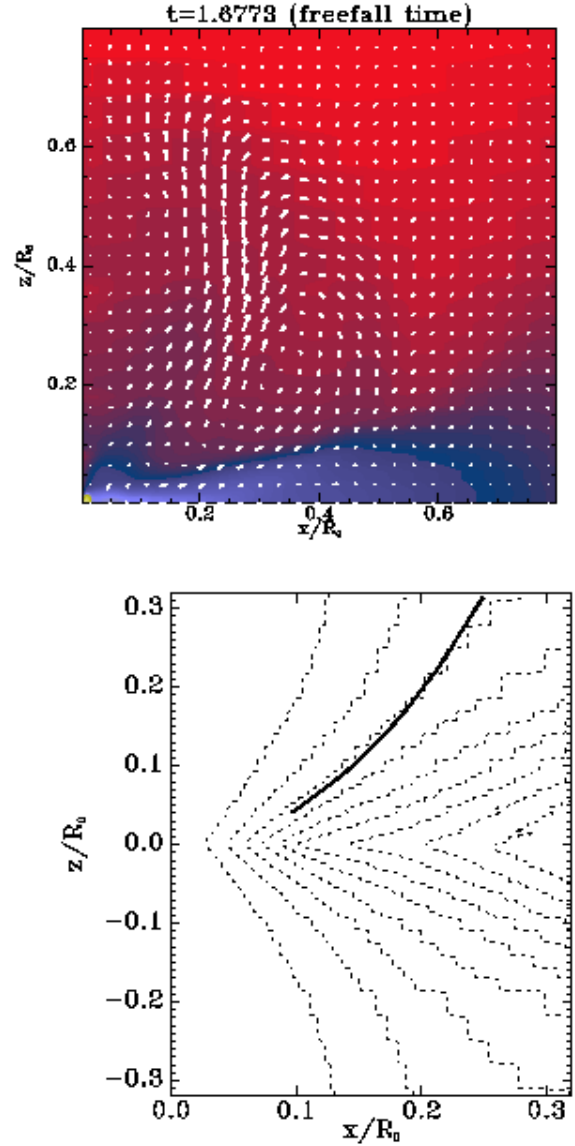}
\caption{ Upper panel: density and velocity field 
 in the model having $\mu=2$ and a critical
  density  $\rho_c / 10$ at time $t=1.67 \tau_{ff}$. 
  Lower panel: structure of the azimuthally averaged poloidal magnetic field
  lines ({\it dashed line}) in the region of the outflow. 
   The  dashed lines show the global structure of 
   those lines, while the
  thick line mark the selected magnetic field line along which some
  quantities will be plotted in Fig.~\ref{angle&force}. Note
  the different scale of the plot compared to
  Fig.~\ref{strongfield_blines}.}
\label{selected_line}
\end{center}
\end{figure}

\begin{figure}
\begin{center}
\includegraphics[width=8cm]{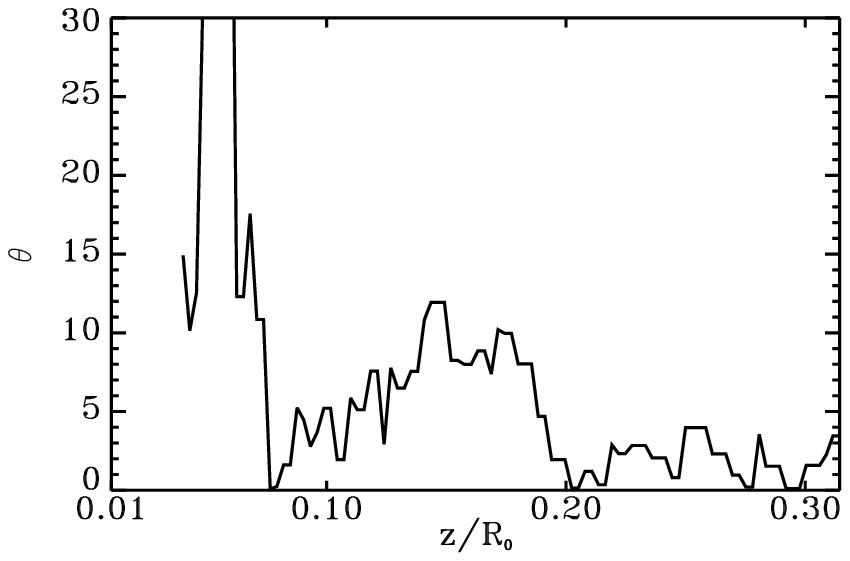}
\includegraphics[width=8cm]{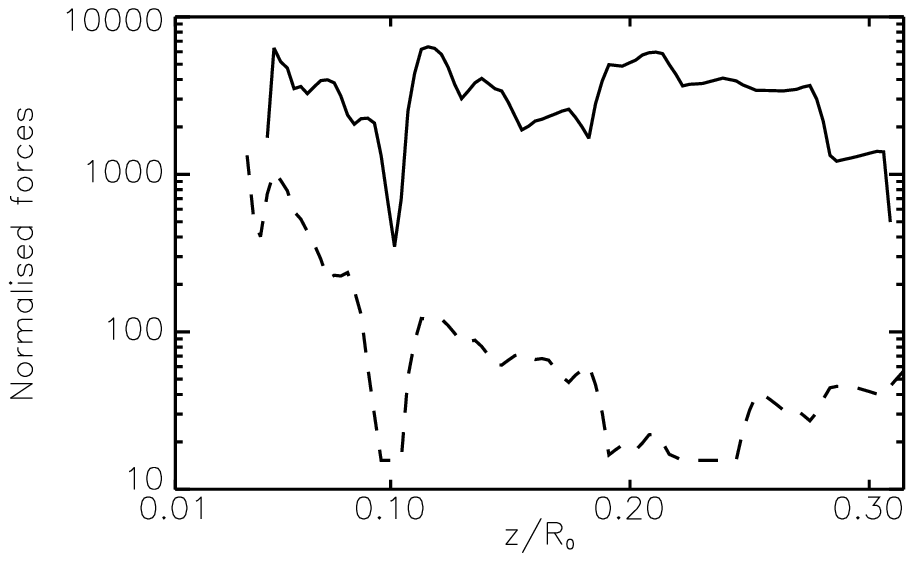}
\includegraphics[width=8cm]{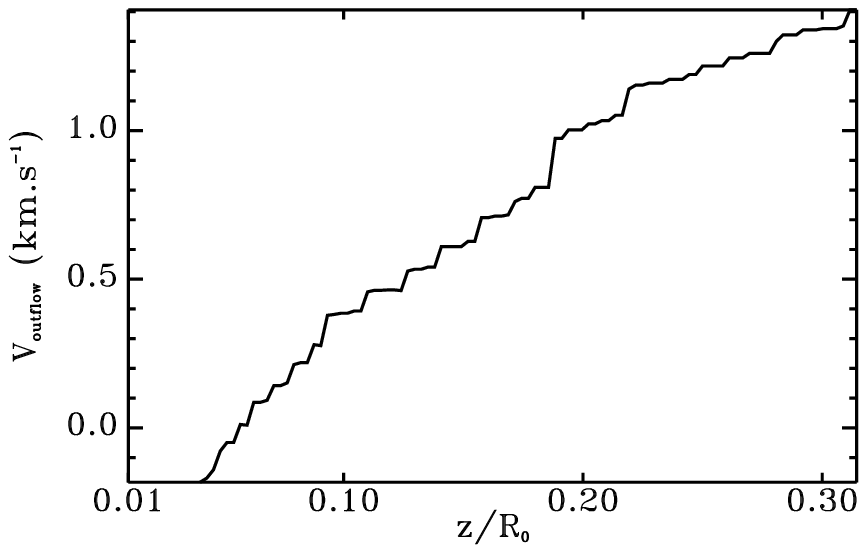}
\caption{First panel: angle $\theta$ (in degrees) between the poloidal fluid
  velocity and the poloidal magnetic field along the particular
  poloidal field line shown in figure~\ref{selected_line}. Note that
  $\theta$ is smaller than $10$ degrees, showing a good alignment
  between both vector. Second panel: Lorentz force ({\it solid line})
  and pressure force ({\it dashed line}) exerted on the fluid element
  along the same magnetic field line. At all positions, the former
  exceeds the latter by one or two orders of magnitude, indicating
  that the outflow is largely magnetically driven.
Third panel: Fluid velocity along the magnetic field line shown in
  Fig.~\ref{selected_line}. The gas is seen to be constantly
  accelerated because of the action of the Lorentz force (see
  fig.~\ref{angle&force}). It reaches a maximum outflow velocity of
  the order of $1.5$ km/s.}
\label{angle&force}
\end{center}
\end{figure}

\begin{figure}
\includegraphics[width=8cm]{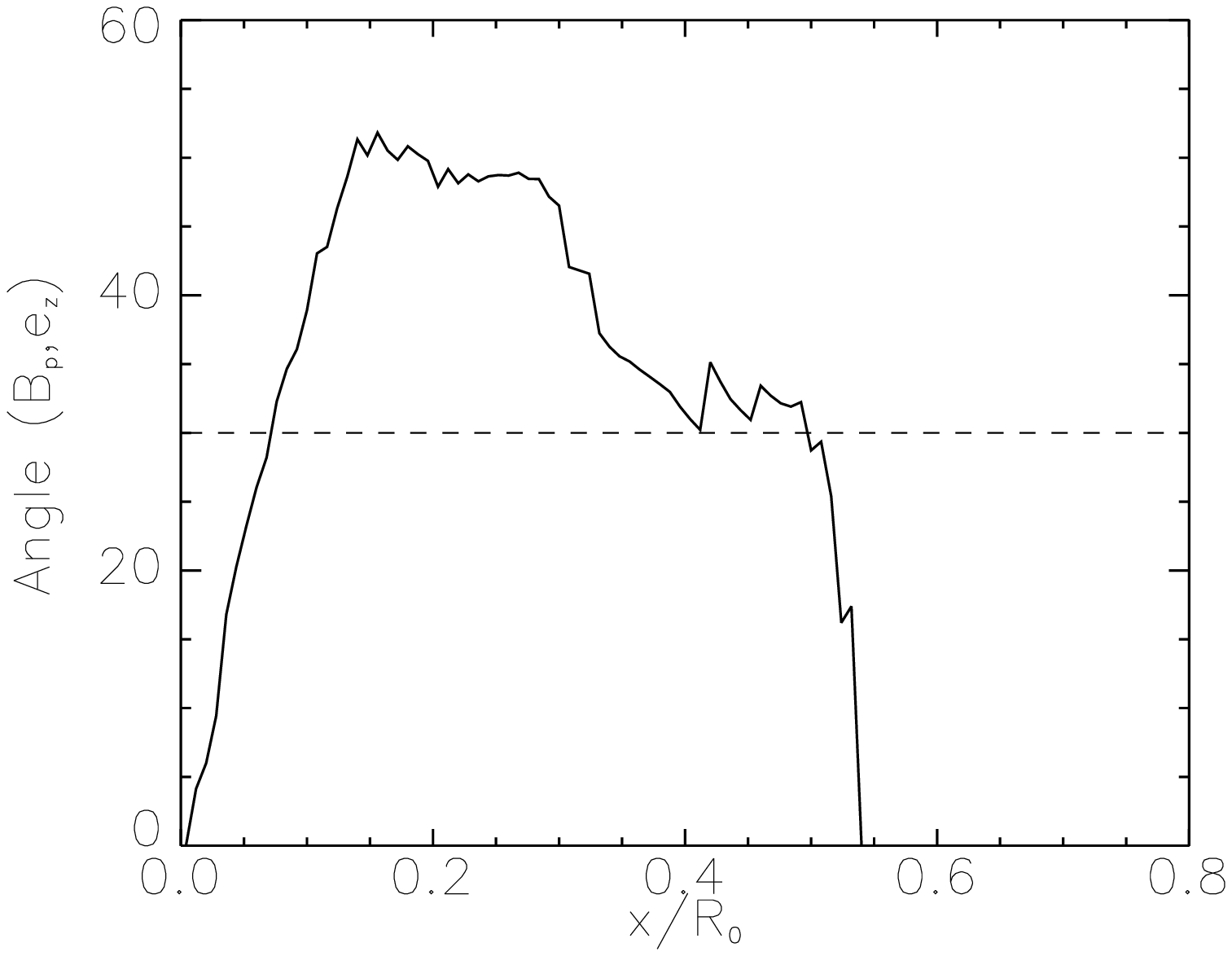}
\caption{Angle between the poloidal magnetic field lines and the disk plane 
close to the disk surface. The analytical theory predicts that this angle should 
be larger than 30 degrees in the outflow launching region.}
\label{bp_angle}
\end{figure}

\subsubsection{Basic features}
In the case $\mu=2$, a collimated outflow developed quickly, as seen
in Fig.~\ref{xz_2}.  The first time displays the density and 
velocity fields just before the outflow is launched.
The second time shows the early phase of the flow
whereas the third time shows a more advanced phase after which 
the flow characteristics do not evolve much (see next section). 
 
The morphology  of the flow is quite 
different from what is obtained in the previous case. 
In particular, there is no slow magnetic tower as in the previous case. This is 
because, as discussed in Sect.~3, there is no centrifugally supported disk, instead
a collapsing magnetic pseudo-disk forms.
The outflows seem to emerge from the central thermally supported core with 
an angle of about $40-45$ degrees with respect to the z-axis and  
 quickly recollimates. Figure~\ref{strongfield_blines}
shows the structure of the magnetic field lines and strength of the 
toroidal magnetic field. 
The poloidal magnetic field is seen to be
mostly vertical, particularly away from the equatorial plane. Close to
the equatorial plane, the magnetic field lines are significantly inclined because
of the inflowing fluid motions.

This is qualitatively in good agreement with the now classical model 
of the magneto-centrifugal ejection  first described by 
Blandford \& Payne (1982) and obtained in many simulations of magnetised disks 
(e.g. Pudritz et al. 2006). In the following section, a more quantitative analysis is presented. 

\subsubsection{Detailed analysis}
The flow features described above tend to suggest that the outflows 
we found in this model are magneto--centrifugally driven. This type of
outflow motion has been studied by many authors using self--similar 
techniques (Blandford \& Payne 1982, Pelletier \& Pudritz 1992) and assuming
stationarity and axisymmetry. Therefore, in order to get  
 the late phase evolution of the outflow for which it is expected that stationarity 
has been reached,  we again 
use the model with a reduced critical density as its dynamical
evolution is faster (larger timesteps in the thermally supported core), 
therefore allowing to get more easily the stationary regime.
All figures were obtained after making an azimuthal average of the
variables around the vertical axis. 

The density and velocity fields are shown in the upper panel of
  Fig.~\ref{selected_line}. The flow is similar to that obtained 
with a critical density equal to $\rho_c$.
To study quantitatively various outflows quantities, we focus 
 on the parts of the outflow which are close to the
equatorial plane ($z/R_0 \le 0.32$). Further away from the disk
midplane, the outflow hits the inflowing material and its structure is
perturbed. The poloidal magnetic field lines in this inner region are
represented in Fig.~\ref{selected_line} with dotted lines. 
The outflow properties are computed along one such field line, represented using the
thick solid line in Fig.~\ref{selected_line}. One of the predictions of
the theories mentioned above is that poloidal velocities $\bf{v_p}$
and magnetic field $\bf{B_p}$ are aligned when the outflow is in steady
state. We plot in the upper panel of Fig.~\ref{angle&force} the
variation of the angle $\theta$ they make as one moves along that
selected field line. Apart from the very inner part of the outflow ($z \le
0.08$, which corresponds to the outflow launching region), $\theta$ is
everywhere smaller than $10$ degrees, indicating a good alignment
between the velocity and the magnetic field. In general, over the
entire outflow region, we found that this angle is always smaller than
$25$ degrees. This is a good indication that the outflow has come close to
reaching steady state, which is in agreement with visual inspections of
animations of this simulation. The middle panel of
Fig.~\ref{angle&force} gives an insight into the launching
mechanism, by plotting the profile of the forces acting on the fluid
along the same field line. The solid line shows the variation of the
component of the Lorentz force along that field line. It is compared with
the pressure gradient, also projected in the same direction. The
former is clearly larger than the later, by one or two orders of
magnitude: the outflow is magnetically (as opposed to thermally)
driven. Finally, we also give the profile of the outflowing velocity along
the magnetic field line (bottom plot of fig.~\ref{angle&force}). Because of the
magnetic force, it increases steadily in the outflow to reach values of
the order of $1.5$ km/s.

Another important prediction of the analytical self-similar model
(Blandford \& Payne 1982) is that the angle between the magnetic field
lines close to the disk and the  z-axis, should be larger than 30
degrees. In Fig.~\ref{bp_angle}, we show this angle as a function of
the radius. It has been measured at the disk surface, defined at each
radius as being the altitude at which the radial fluid velocity
vanishes. It is seen that this angle is indeed always larger than 30
degrees except in the very center and in the outer part. In these two
regions, no outflow is occurring as can be seen in Fig.~\ref{selected_line}. 


\begin{figure}
\includegraphics[width=8cm]{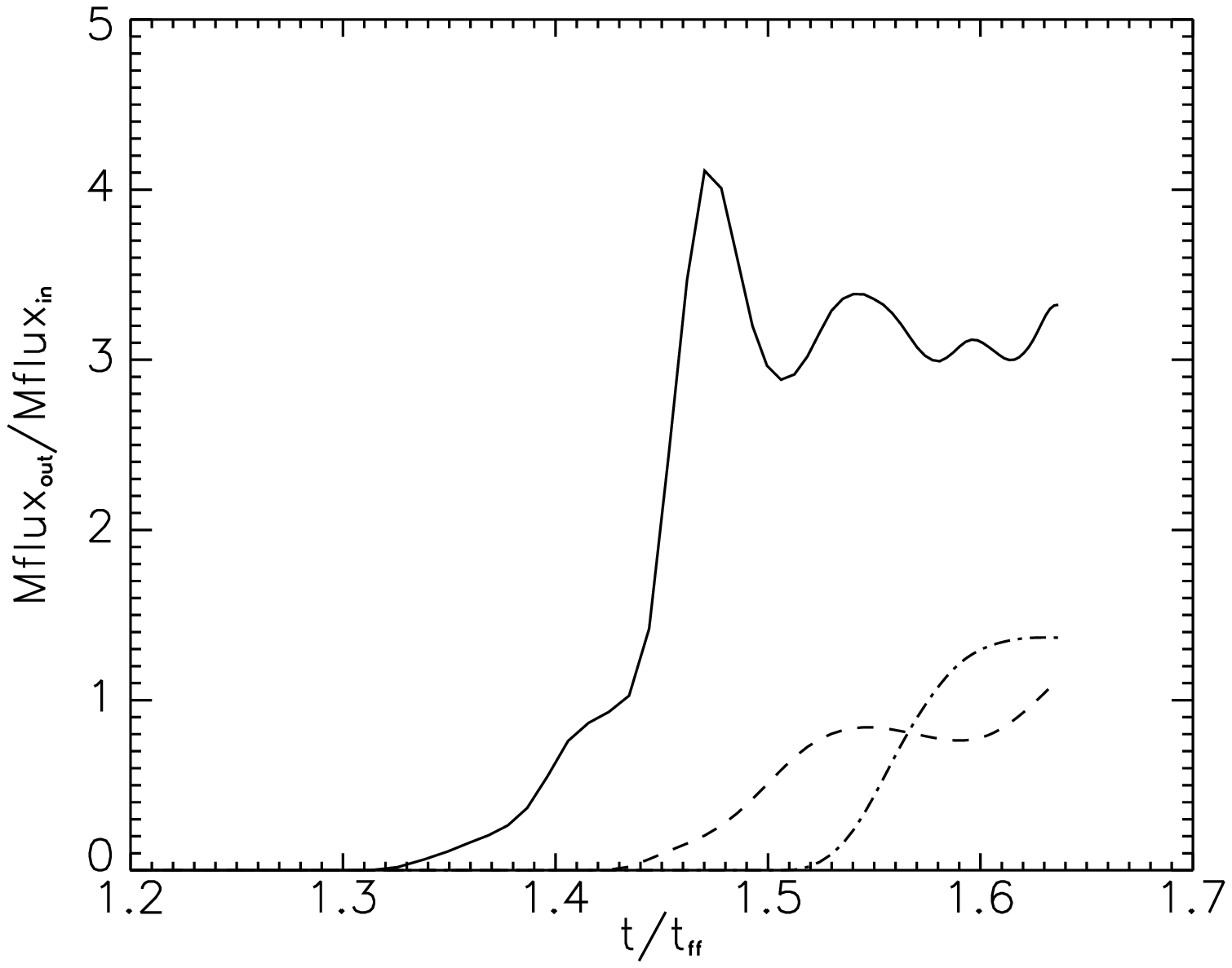}
\includegraphics[width=8cm]{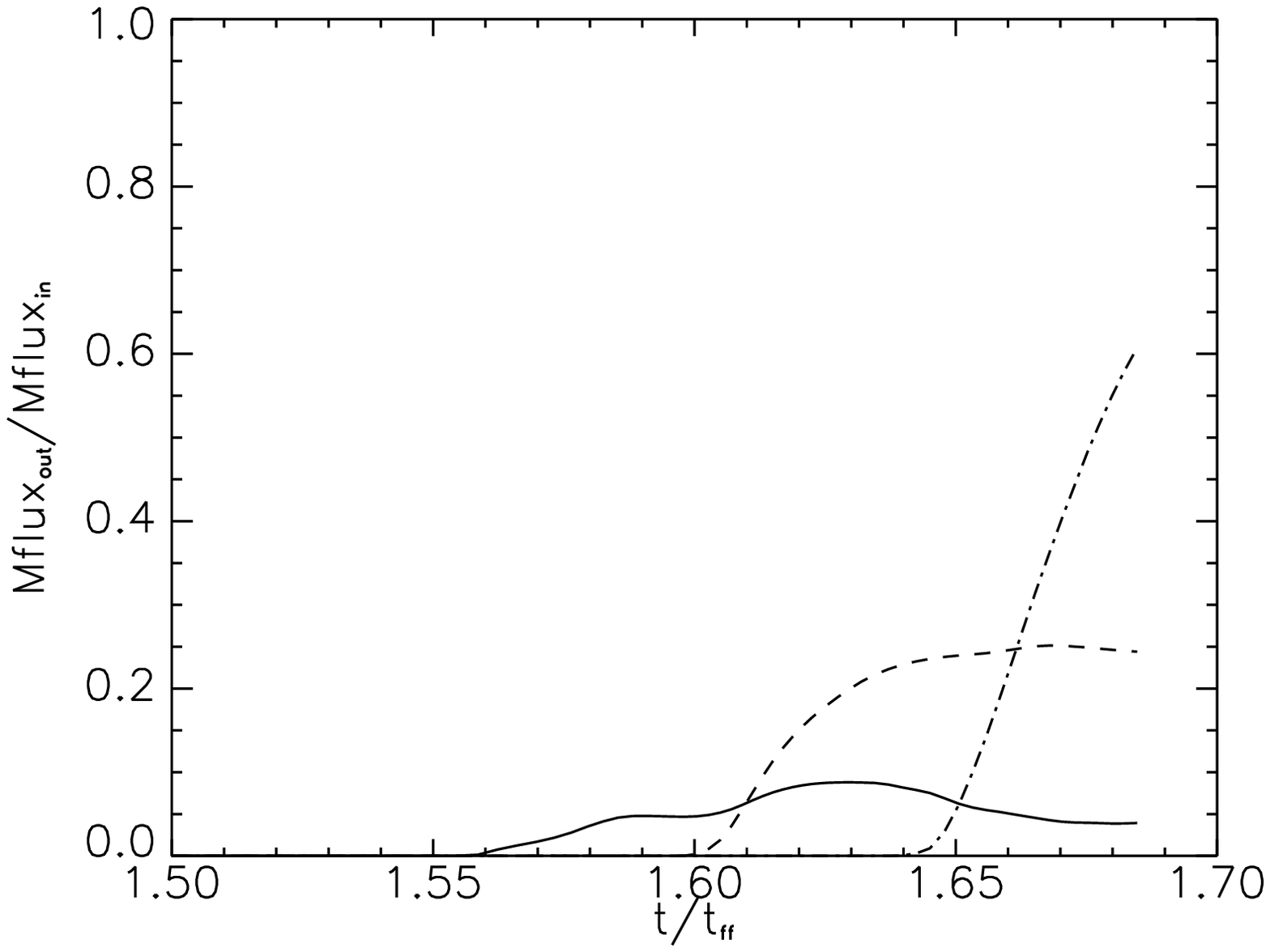}
\caption{Ratio of ejection over accretion  mass rate as a 
function of time. Upper panel is $\mu=20$ and lower panel $\mu=2$.
Accretion and ejection mass rates are estimated on a sphere 
of radius $R_s$.
Solid lines are for $R_s/R_0=0.2$, dotted lines for $R_s/R_0=0.4$ whereas 
dashed lines are for $R_s/R_0=0.6$.}
\label{massflux}
\end{figure}

\begin{figure}
\includegraphics[width=8cm]{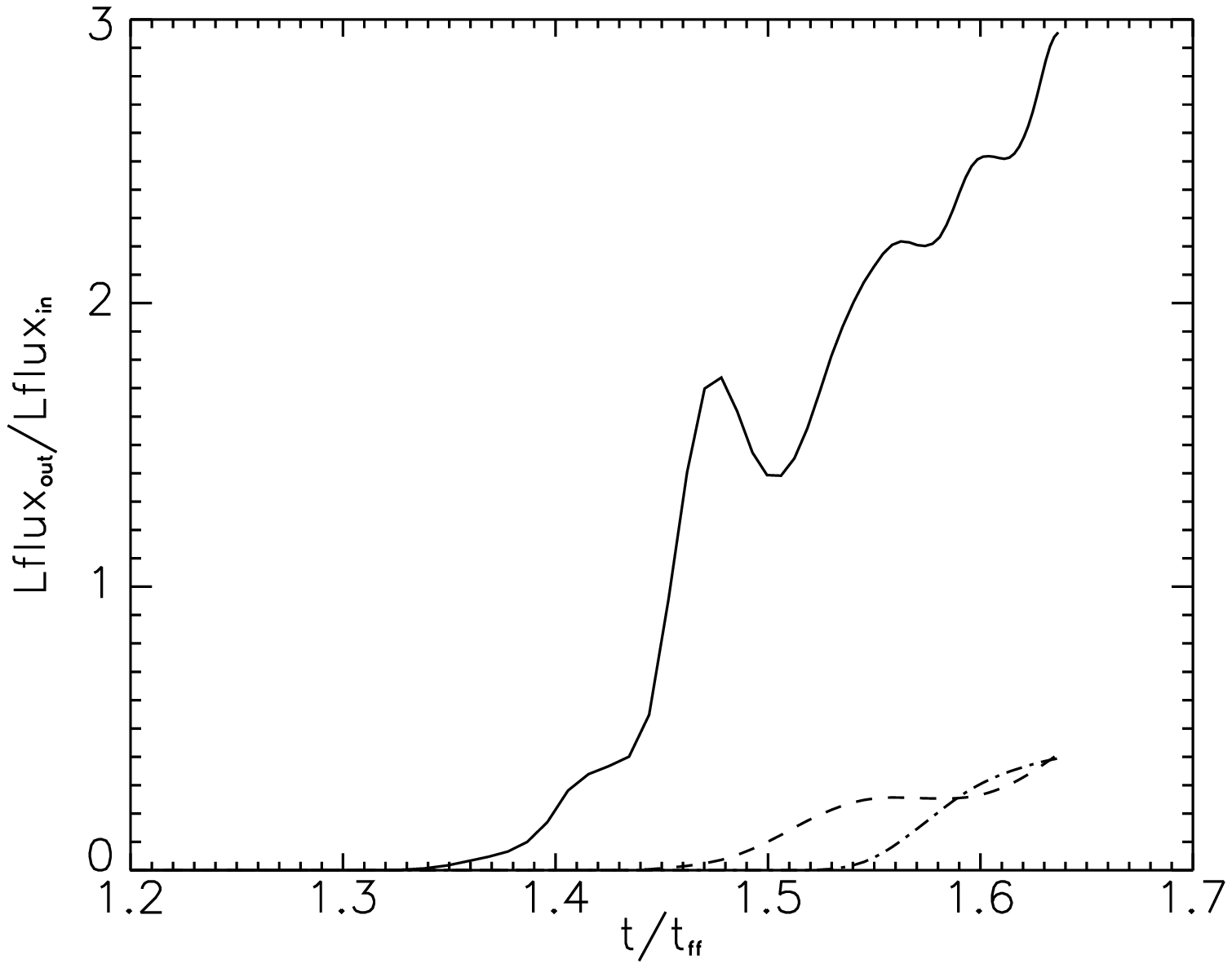}
\includegraphics[width=8cm]{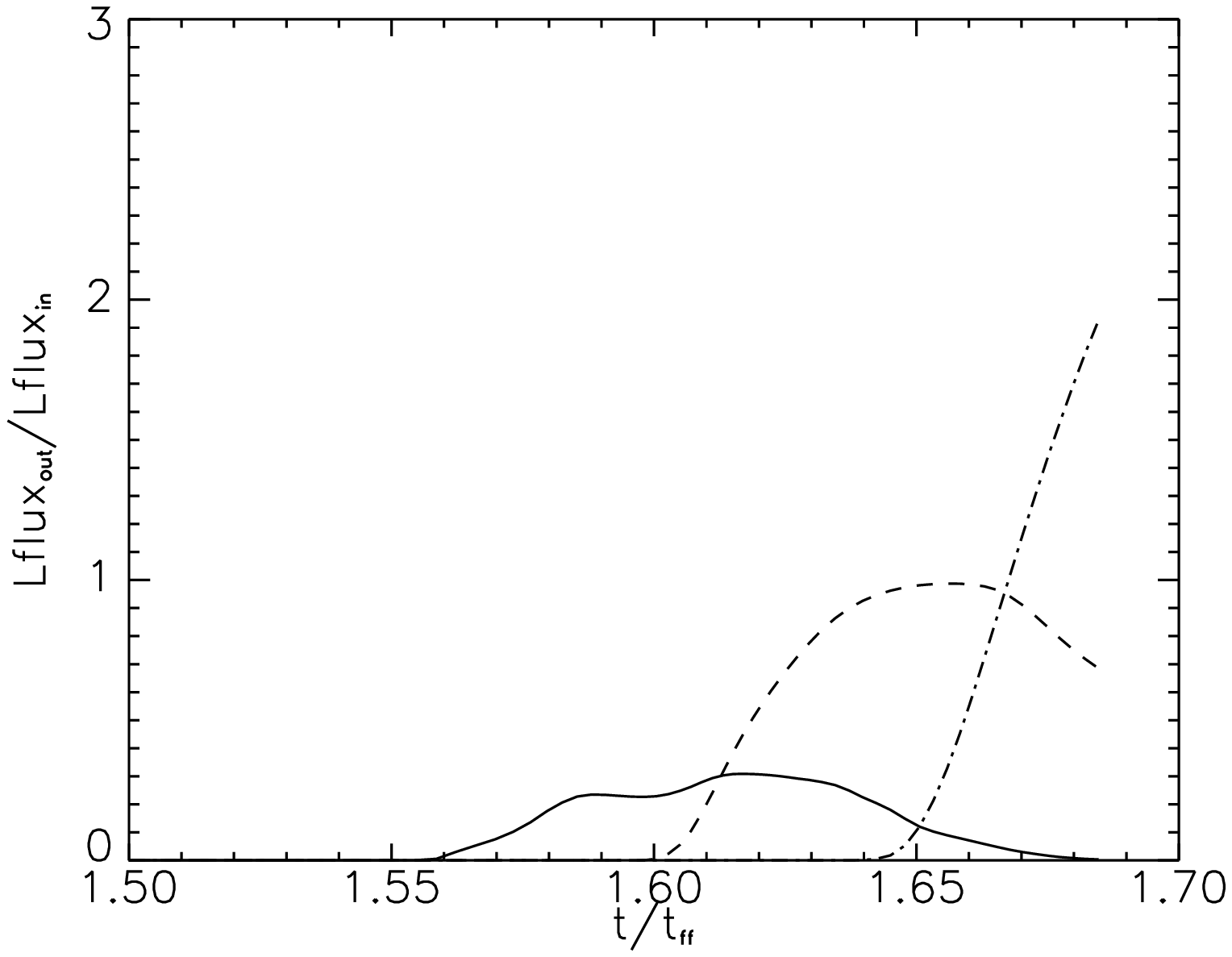}
\caption{Ratio of  ejection over accretion angular momentum rate as a 
function of time. Upper panel is $\mu=20$ and lower panel $\mu=2$.
Accretion and ejection angular momentum rates are estimated on a sphere 
of radius $R_s$.
Solid lines are for $R_s/R_0=0.2$, dotted lines for $R_s/R_0=0.4$ whereas 
dashed lines are for $R_s/R_0=0.6$.}
\label{momflux}
\end{figure}

\subsection{Mass and angular momentum fluxes}
We now present quantities which characterize globally the 
evolution of the whole accretion-ejection structure with time. For this purpose 
we again use the simulations with the critical density, $\rho_c/10$, 
 since they allow to follow the cloud evolution further. 

Figures~\ref{massflux} and~\ref{momflux} display the 
ratio of  ejected over accreted mass and angular momentum fluxes. 
They are estimated on spheres of various radius $R_s$, namely 
$R_s/R_0=0.2$ (solid lines), $R_s/R_0=0.4$ (dotted lines) and 
$R_s/R_0=0.6$ (dashed lines). Note that for $\mu=20$, the first value
of $R_s$ is inside the magnetic tower whereas the 2 other values
correspond to radius higher than the equatorial radius of the magnetic
tower.

For $\mu=20$ and $R_s/R_0=0.2$,  the ratio of ejection over accretion
mass rate vanishes 
before $t=1.3 \tau_{ff}$, then increases until a value of about 3-4. At this 
point  quasi-stationarity is reached.
This indicates  that because of the centrifugal barrier, 
the gas first piles up in the inner region. Then the magnetic tower and the outflow
described previously extract efficiently the mass at a rate higher than the 
accretion mass rate.  For the two larger values of $R_s$, the 
ratio of ejection over accretion mass rate is smaller by a factor of about 3.
This is due to  higher accretion rates in the collapsing envelope than in the 
inner centrifugally supported structure.

The behaviour for $\mu=2$, is much different. The 
ratio of ejected over accreted mass rate 
varies between 0.1 and 0.6 and is therefore always smaller than 1.
 It increases with $R_s$. In that case, the gas falls directly in the centre 
without piling up in a centrifugally supported structure. The
saturated ratios obtained for the two smallest $R_s$ are reminiscent
of typical values quoted in the literature. 

The ratio of ejection over accretion angular momentum rates  have a
similar behaviour than the ratio of ejection over accretion mass rates. However, 
we note that for $\mu=20$, the former is smaller than the later by a factor 
1 to 2 whereas for $\mu=2$, the contrary is true (the ratio being as high as 3). 
These differences are due to the fact that in case $\mu=20$, the transportation of 
the angular momentum is weak since the magnetic tension is weak. Therefore the 
angular momentum is not transferred efficiently and is mostly advected with the gas.
Since the gas which is accreted comes from larger radius than  the gas which is ejected, 
 the later has on average a larger angular momentum than the former. 
On the contrary, in case $\mu=2$, the gas is efficiently braked near the equatorial plane 
whereas it is  azimuthally accelerated at higher altitude therefore carrying with it 
a higher angular momentum. 

\section{Comparison with observations}
Here we qualitatively discuss comparisons between the models presented 
in the previous sections and various observations. 
One of the difficulties in carrying out detailed comparisons between 
observations and models of  protostellar dense core is the need for sources  
sufficiently constrained observationally.

\subsection{The case of IRAM04191}
In this respect, the 1.5 solar mass, young class 0 source IRAM04191 (Andr\'e et al. 1999, 
Belloche et al. 2002) located in the Taurus molecular cloud,  is  a nice 
example. In this elongated source, an outflow perpendicular to the major axis 
of the source
has been observed suggesting that the rotation axis is also perpendicular
to the major axis. With these assumptions, the rotation velocity 
has been measured.
Moreover, the radial  velocities and  the column density profile are known as well
 from radiative modeling of the line profiles.
A dynamical age of $\simeq$2 10$^4$ years has  been estimated from the characteristic
of the flow. Finally, no disk has been detected in this source, the upper limit
for the disk radius being around 15 AU.

Various attempts have been made to compare these profiles with hydrodynamical models, 
starting initially with 
critical Bonnor-Ebert sphere in rotation (Belloche 2002, Hennebelle et al. 2004b, Lesaffre et al. 2005). 
These models fail to reproduce IRAM04191 for the following reasons.  First, the infall velocity 
($\simeq$ 0.15 km/s)
is too large at $r\simeq 2000$ AU in the model (0.2-0.3 km/s). 
Second the column density in the inner part ($r< 1000$ AU) is too large in the model. 
In the model, the large column density in the inner region
 is due to the influence of rotation (see Fig.~\ref{mu=1000}) which provides a support to the 
infalling gas.
Self-consistently, the rotation curve cannot be reproduced under simple assumptions
on the initial angular momentum distribution. In particular, the rotation velocity of the hydrodynamical 
model  tends to be  too high in the 
inner part. Another important related disagreement, is the absence of a big massive disk 
in IRAM04191 like the one produced in the hydrodynamical simulation.

Although no detailed comparison has been carried out, it seems worthwhile to   compare
with the magnetized models presented in this paper qualitatively. 
The first interesting aspect is the infall velocity
which is smaller in the outer part than in the hydrodynamical case because of magnetic support. 
Although in the models presented here, it is still too large to reproduce the infall of 
IRAM04191, starting with near equilibrium configuration
will help to reduce it further. However, running specific models
dedicated to this particular case is beyond the scope
of this paper. 
 The second aspect is the rotation curve, which is much flatter in the 
$\mu=2-5$ cases  than in the hydrodynamical case. This is also in better agreement with the rotation 
curve inferred for IRAM04191.  Finally, the models with $\mu < 5$ do not show the presence of 
hundred AU size disk unlike the hydrodynamical model. It is indeed, extremely difficult to 
reconcile the absence of disk and the presence of rotation within the framework of 
hydrodynamical models. 

Another interesting aspect is the outflow which is observed in IRAM04191 (Andr\'e et al. 1999). 
Its velocity is about 
5-10 km/s. This is qualitatively in good agreement with the outflows spontaneously launched 
in the MHD models. 
Quantitatively, however, this is 2 to 5 times larger than the values obtained in this paper
(note that outflow velocities up to $\simeq$3 km s$^{-1}$ are produced in our simulations).
Nevertheless,  as recalled previously, one should remember
that the speed of the outflow is related to the rotation velocity achieved by the fluid particles. 
Since  the physics of the first Larson core and the second collapse are not treated in this 
work,
 further contraction of the gas is prevented and  therefore the velocity of the outflow is reduced. 
It seems therefore
reasonable to assume that a better treatment of the first Larson core will yield to faster outflows (see 
Banerjee \& Pudritz 2006, Machida et al. 2007).

\subsection{Observations of  class 0 sources with disk}
While disks are commonly observed during the late stage of star formation, i.e. class I, class II or
 T Taury phase (e.g. Watson et al. 2007), disks are more difficult to observe  during the class 0
phase and therefore much less constrained (Mundy et al. 2000).
Nevertheless, various studies report class 0 protostars having disks of masses between 1 and 10 percents 
of their envelope masses (Looney et al. 2003, Jorgensen et al. 2007) giving a typical 
mass of about 0.1 solar mass.

Since the age of these sources is not well known and the density and velocity  profiles are not
available,  it appears difficult to reach solid conclusions. This may
nevertheless indicate that in these sources, the magnetic field is not too strong with typically 
$\mu > 5$.

\section{Conclusion}
Using RAMSES, we  performed 3D simulations of a 
magnetised collapsing dense core. We explored the effect of the initial 
magnetic field strength varying the value of the mass-to-flux over critical 
mass-to-flux ratio, $\mu$, from 1000 to 2. The cloud evolution 
is significantly modified for values as large as $\mu=20$. This is due 
to the strong amplification of the radial and azimuthal components of the 
magnetic field induced by the differential motions arising in the 
collapsing cloud.
 
We also developed semi-analytical models that predict some of the core envelope properties 
and compared them with the simulations results showing reasonable agreements. 

For $\mu=20$, we find that magnetic braking is negligible and that consequently 
a centrifugally supported disk forms. A magnetic  tower, 
generated by the twisting of the field lines, forms and expands reducing the mass 
of the centrifugally supported structures. A faster outflow is then triggered from the 
thermally supported central core. 

For $\mu$ smaller than 5,  no centrifugally supported disk forms for two
reasons. First the collapse occurs primarily along the field lines
which means that less angular momentum is delivered to the inner
parts. Second, strong magnetic braking extracts the angular momentum
from the disk. The question as to whether a disk will  form at later stage 
remains nevertheless open. In addition, 
an outflow is  triggered from the thermally supported core in that case. 
Detailed investigations have been performed in the case $\mu=2$. 
They reveal that the outflow reaches a quasi-steady state 
regime and features many characteristics of the magneto-centrifugally
driven outflow models studied analytically  in the literature (e.g. Blandford \& Payne 1982, 
Pudritz et al. 2006).

In a companion paper, we study the fragmentation of the collapsing dense core paying particular
attention to the strength of the magnetic field. The analysis developed in the
 present paper
is then used to interpret the numerical results obtained in the context of fragmentation.

\acknowledgement
 Some of the simulations
presented in this paper were performed at the IDRIS supercomputing center and on the CEMAG 
computing facility supported by the frensch ministry of research and education through 
 a Chaire d'Excellence awarded to Steven Balbus.
We thank Romain Teyssier, Doug Johnstone and Philippe Andr\'e for  critical reviews
 of the manuscript as well as  Frank Shu, the referee for helpful comments.
PH acknowledge many discussions on related topics over the years with Daniele Galli and discussions
with Zhi-Yun Li about the recent work of Mellon \& Li. He also thanks Sylvie Cabrit,
Fabian Casse and Jonathan Ferreira for discussions about the physics of outflows.

\appendix

\section{An analytical model for the expansion of the magnetic tower with self-gravity}

\begin{figure}
\includegraphics[width=7cm]{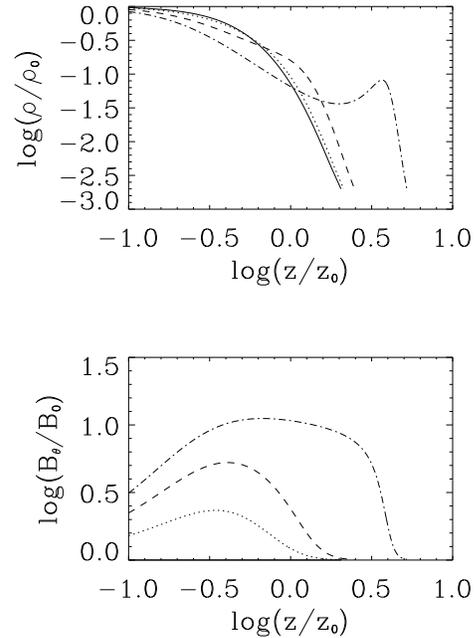}
\caption{Density and transverse component of the magnetic field as given 
by Eqs.~(\ref{sol_B}) and~(\ref{rho}) for four times as a function of $z$.}
\label{bulana}
\end{figure}

We present an analytical model to investigate the mechanism responsible for the 
expansion of the self-gravitating disk. 

In the simulation, the problem appears to be axisymmetric making 
it bidimentional. It is also evidently time dependent. 
For the purpose of reducing the complexity, 
we consider  a self-gravitating layer  which varies along the z-axis only,
instead of an axisymmetric structure. We therefore replace, 
the azimuthal fields, $B_\theta$ and $V_\theta$ by the plane-parallel transverse 
fields $B_x(z)$ and $V_x(z)$.
 It is initially 
threaded by a vertical and uniform magnetic field, $B_z^0$. 
The transverse velocity, $V_x(z)$, initially generates a transverse 
magnetic field, $B_x(z)$ which modifies the vertical equilibrium.
By doing so, we ignore the effect of the curvature inherent to the 
axisymmetric structure.

For simplicity, we  make the approximation that the layer is in mechanical 
equilibrium and use Eq.~(\ref{z_equilibrium}) written previously.
Let $\sigma= 2 \int \rho dz$ be the column density.
With the Poisson equation, we get 
\begin{eqnarray}
\partial_z \psi = -  2 \pi G \sigma 
\label{poisson}
\end{eqnarray}

Thus, defining 
\begin{eqnarray}
\widetilde{\rho} = \rho / \rho_0 \,  , \, 
\widetilde{\sigma } = \sigma / \sqrt{2 \rho_0 c_s^2 /  \pi G} \, , \, 
\widetilde{B}_x = B _x / B_0, 
\end{eqnarray}
where $B_0 = \sqrt{4 \pi \rho_0 c_s^2}$, 
 we get, since $\sigma(0)=0$ and $B_x(0)=0$,
\begin{eqnarray}
    \widetilde{\rho} +   \widetilde{\sigma}^2  +
{ \widetilde{B_x} ^2 \over 2 }  =  1,
\label{mouv_z}
\end{eqnarray}
where $\rho_0$ is the density at $z=0$.

In order to compute $B_x$, we write the transverse momentum and induction 
equations using the Lagrangian coordinates (Shu 1983). 
With $\widetilde{t} = t / t_0$ where $t_0 = 1 / \sqrt{2 \pi \rho_0 G} $,
and $\widetilde{z} = z / z_0$ where $z_0 = \sigma_0 / 2 \rho_0$ we have
\begin{eqnarray}
 d_{\widetilde{t}} \widetilde{V}_x  = 
\widetilde{B}_z^0 \partial_{\widetilde{\sigma}} \widetilde{B}_x, 
\label{mouv_thet}
\end{eqnarray}
and
\begin{eqnarray}
d_{\widetilde{t}} \left( { \widetilde{B}_x \over \widetilde{\rho} } \right)  = 
\widetilde{B}_z^0 \partial_{\widetilde{\sigma}} \widetilde{V}_x
\label{ind_thet}
\end{eqnarray}
This equation shows that, in the model, the growth of the toroidal field is triggered 
only by the vertical gradient of $V_x$, the transverse velocity.
It is worth stressing that this equation does not include all the physical relevant 
terms in the problem. In particular, in the simulation the growth of the toroidal field is largely 
due to the twisting of the radial component of the magnetic field by the differential 
rotation which cannot be included in a plane parallel geometry. 
We note however that Fig.~\ref{magnetic tower}  clearly shows vertical gradients 
of $V_\theta$ ($z>10^{-4}$ pc at time 1.116 and 1.125 $\tau_{ff}$).

Equations~(\ref{mouv_z}),~(\ref{mouv_thet}) and~(\ref{ind_thet}) are to be solved.
Despite the simplifications, i.e. dependence on $z$ only and mechanical
equilibrium in the vertical direction, they are still complex two variable
non-linear equations. 
Since the disk is symmetric with respect to the equatorial plane, 
the boundary conditions are $B_x(0) = 0$ and $\partial_\sigma V_x (0) = 0$.

In order to illustrate the origin of the expansion due to the growth of the 
toroidal magnetic component, we consider as initial conditions a self-gravitating
layer with a vanishing $B_x$ at time $\widetilde{t}=0$. In that case the solution is simply 
$\widetilde{\rho} = 1 - \widetilde{\sigma}^2$.  
In term of the $\widetilde{z}$ variable, this writes $\widetilde{\rho}(\widetilde{z}) = 1 / {\rm ch}(2\widetilde{z})^2$
(Spitzer 1942, Ledoux 1951, Curry 2000). 

Since obtaining exact solutions of Eqs.~(\ref{mouv_z})-(\ref{ind_thet}) seems to be difficult, 
we seek   approximated
 solutions of the equations written above. To this purpose, we replace 
Eq.~(\ref{ind_thet}) by
\begin{eqnarray}
d _{\widetilde{t}} \widetilde{B}_x  = 
\left(1 - \widetilde{\sigma}^2 \right)  
\partial_{\widetilde{\sigma}} \widetilde{V}_x,
\label{ind_thet_approx}
\end{eqnarray}
that is to say, we assume that the density in Eq.~(\ref{ind_thet}), is the density of 
the unmagnetised solution. Strictly speaking, this approximation holds as long as 
the density has not been modified  by the magnetic field significantly. 
With this assumption, the time and space variables separate, making it easy 
to find solutions as
\begin{eqnarray}
\widetilde{V}_x(\widetilde{t},\widetilde{\sigma})= 
\widetilde{V}_0 \cos(\sqrt{6} B_z^0 \widetilde{t}) \times (1 - 3 \widetilde{\sigma}^2), 
\label{sol_V}
\end{eqnarray}
\begin{eqnarray}
\widetilde{B}_x(\widetilde{t},\widetilde{\sigma})= -\sqrt{6} \widetilde{V}_0 \sin(\sqrt{6}
 \widetilde{B}_z^0 \widetilde{t}) \times 
(\widetilde{\sigma} -  \widetilde{\sigma}^3).
\label{sol_B}
\end{eqnarray}
Using Eq.~(\ref{mouv_z}), we obtain the density as a function of $\widetilde{\sigma}$ and 
$\widetilde{t}$
\begin{eqnarray}
\widetilde{\rho} = 1 - \widetilde{\sigma}^2 - 3 \widetilde{V_0}^2 \sin^2(\sqrt{6} 
\widetilde{B}_z^0 \widetilde{t})
(\widetilde{\sigma}  -\widetilde{\sigma}^3)^2.
\label{rho}
\end{eqnarray}
using the relation $d \sigma = 2 \rho dz$, it is possible to obtain $z$ as a function 
of $\sigma$ and $\rho$. 

Figure~\ref{bulana} shows $\widetilde{\rho}$ and $\widetilde{B}_x$ as given by 
Eqs.~(\ref{sol_B}) and~(\ref{rho}) at four times for $\widetilde{V}_0=2$
and $\widetilde{B}_z^0=1$.
The behaviour is very similar to the evolution displayed in Fig.~\ref{magnetic tower}.
As $\widetilde{B}_x$ is growing, the layer is expanding because of the magnetic pressure. 

\end{document}